\documentclass[12pt]{article}
\usepackage{epsfig, epsf, graphicx, subfigure}
\usepackage{pstricks, pst-node, psfrag}
\usepackage{amssymb,amsmath,bm}
\usepackage{verbatim,enumerate}
\usepackage{rotating, lscape}
\usepackage{setspace}
\usepackage[english]{babel}
\usepackage[compress]{natbib}
\usepackage{multirow}
\usepackage{theorem}
\usepackage{array}

\def\d{\mathrm{d}}

\def\00{\mathrm{0}}

\def\ZZ{\mathbf{Z}}

\def\CC{\mathcal{C}}

\def\AA{\mathcal{A}}

\def\zz{\boldsymbol{z}}
\def\tht{\boldsymbol{\theta}}

\def\ee{\mathbf{e}}
\def\uu{\mathbf{u}}
\def\yy{\mathbf{y}}

\def\ii{\boldsymbol{1}}

\def\ss{\boldsymbol{s}}

\newtheorem{prop}{Proposition}
\newtheorem{corol}{Corollary}
\newtheorem{example}{Example}
\newtheorem{lemma}{Lemma}

\begin{document}
	
\thispagestyle{empty} \baselineskip=28pt \vskip 5mm
\begin{center} 
%{\Huge{\bf Cauchy convolution processes for the modeling of spatial extremes with local tail dependence}}
%{\Huge{\bf Cauchy convolution processes for modeling spatial data with local tail dependence}}
%{\Huge{\bf Cauchy convolution processes to model spatial data with short-range tail dependence}}
%{\Huge{\bf Spatial Cauchy convolution processes for modeling short-range tail dependence}}
%{\Huge{\bf Modeling Spatial Data with Cauchy Convolution Processes}}
{\Huge{\bf Modeling Spatial Tail Dependence with Cauchy Convolution Processes}}
%{\Huge{\bf Cauchy Convolution Processes to Model Spatial Data}}
\end{center}
	
\baselineskip=12pt \vskip 10mm
	
\begin{center}\large
		Pavel Krupskii\footnote[1]{\baselineskip=10pt University of Melbourne, Parkville, Victoria, 3010, Australia. E-mail: pavel.krupskiy@unimelb.edu.au.}, 
		Rapha\"{e}l Huser\footnote[2]{\baselineskip=10pt Statistics Program, Computer, Electrical and Mathematical Sciences and Engineering (CEMSE) Division, King Abdullah University of Science and Technology (KAUST), Thuwal 23955-6900, Saudi Arabia.. E-mail: raphael.huser@kaust.edu.sa.}
\end{center}
	
\baselineskip=17pt \vskip 10mm \centerline{\today} \vskip 15mm
	
%%%%%%%%%%%%%%%%%%%%%%%%%%%%%%%%%%%%%%%%%%%%%%%%%%%%%%%%%%%%%%%%%%%%%%%%
\begin{center}
	{\large{\bf Abstract}}
\end{center}
	
We study the class of dependence models for spatial data obtained from Cauchy convolution processes based on different types of kernel functions. We show that the resulting spatial processes have appealing tail dependence properties, such as tail dependence at short distances and independence at long distances with suitable kernel functions. We derive the extreme-value limits of these processes, study their smoothness properties, and detail some interesting special cases. To get higher flexibility at sub-asymptotic levels and separately control the bulk and the tail dependence properties, we further propose spatial models constructed by mixing a Cauchy convolution process with a Gaussian process. We demonstrate that this framework indeed provides a rich class of models for the joint modeling of the bulk and the tail behaviors. Our proposed inference approach relies on matching model-based and empirical summary statistics, and an extensive simulation study shows that it yields accurate estimates. We demonstrate our new methodology by application to a temperature dataset measured at 97 monitoring stations in the state of Oklahoma, US. Our results indicate that our proposed model provides a very good fit to the data, and that it captures both the bulk and the tail dependence structures accurately.
	
\baselineskip=14pt

\par\vfill\noindent
{\bf Keywords:}
Copula; Extreme-value model; Kernel convolution process; Short-range spatial dependence; Spatial process; Tail dependence
%\par\medskip\noindent
%{\bf Short title}: Spatial processes with local tail dependence
	
\clearpage\pagebreak\newpage \pagenumbering{arabic}
\baselineskip=25pt

\allowdisplaybreaks

%%%%%%%%%%%%%%%%%%%%%%%%%%%%%%%%%%%%%%%%%%%%%%%%
%%%%%%%%%%%%%%%%%%%%%%%%%%%%%%%%%%%%%%%%%%%%%%%%
%%%%%%%%%%%%%%%%%%%%%%%%%%%%%%%%%%%%%%%%%%%%%%%%
\section{Introduction}\label{sec:intro}

{Assessment of environmental risk associated with unprecedented air or sea temperatures \citep{Davison.Gholamrezaee:2012,Huser.Genton:2016,Hazra.Huser:2020}, extreme flooding \citep{Thibaud.etal:2013,Huser.Davison:2014a,CastroCamilo.Huser:2020}, strong wind gusts \citep{Engelke.etal:2015,Oesting.etal:2017,Huser.etal:2017}, or high air pollution levels \citep{Vettori.etal:2019,Vettori.etal:2020} requires the computation of joint tail probabilities. Because the process of interest is always observed at a finite set of monitoring sites, spatial modeling is needed whenever the required probabilities involve one or more unobserved locations, and the assumed tail dependence structure plays a crucial role in estimating these risks.}

{Gaussian processes have been widely used in the literature to model spatio-temporal dependence, because they are computationally convenient and they are parameterized using various types of covariance functions that can capture features such as the dependence range and the smoothness of realizations (see, e.g., \citet{Gneiting2002} and \citet{Gneiting.Genton.ea2007}). However, Gaussian models have restrictive symmetries and cannot capture strong tail dependence that is often found in spatial data, which makes them unsuitable when the main interest lies in the tails. More flexible yet computationally feasible spatial models are required. To circumvent the limitations of multivariate normality, copula models have become increasingly popular and have found a wide range of environmental applications in geology \citep{Graler.Pebesma2011}, hydrology \citep{Bardossy2006, Bardossy.Li2008} and climatology \citep{Erhardt.Czado.ea2014}, among others. A copula is simply defined as a multivariate cumulative distribution function (CDF) with standard uniform ${\rm Unif}(0,1)$ marginals. \citet{Sklar1959} showed that for any continuous $d$-dimensional CDF $F$ with marginals $F_1, \ldots, F_d$ there exists a unique copula $C$ such that $F(z_1,\ldots,z_d) = C\{F_1(z_1), \ldots, F_d(z_d)\}$. A random vector $(Z_1,Z_2)^\top$ with margins $F_1,F_2$ and copula $C$ is said to be upper tail-dependent if the limit
\begin{equation}\label{eq:lambdaU}
\lambda_U=\lim_{u\to1}\Pr\{Z_1>F_1^{-1}(u)\mid Z_2>F_2^{-1}(u)\}=\lim_{u\to1}{1-2u+C(u,u)\over1-u},
\end{equation}
exists and is positive, i.e., $\lambda_U>0$, and is upper tail-independent if $\lambda_U=0$. An analogous definition holds for the lower tail based on the coefficient $\lambda_L=\lim_{u\to0}\Pr\{Z_1\leq F_1^{-1}(u)\mid Z_2\leq F_2^{-1}(u)\}=\lim_{u\to0}{C(u,u)/u}$. A two-dimensional copula $C$ is said to be tail-symmetric if $\lambda_L=\lambda_U$ and is reflection-symmetric if $C=C^R$, where $C^R(u_1,u_2):=-1+u_1+u_2+C(1-u_1,1-u_2)$ is the reflected copula. While multivariate normal vectors are always reflection-symmetric, as well as tail-independent when the correlation is less than one \citep{Ledford.Tawn:1996}, other copula models can be used to construct flexible multivariate distributions with arbitrary marginals and various tail dependence structures. Certain copula families, such as vine models \citep{Aas.Czado.ea2009,Kurowicka.Joe2011}, are very flexible but lack interpretability with spatial data. By contrast, factor copula models \citep{Krupskii.Joe:2013} have been proposed as flexible models capturing non-Gaussian features like reflection and/or tail asymmetry and strong tail dependence, and they can be naturally extended to the spatial context; see \citet{Krupskii.Huser.ea2016} and \citet{Krupskii.Genton2016}. However, since these models are built from common underlying random factors affecting all spatial sites simultaneously, they are unable to capture full independence at large distances. Hence, these models are only suitable for spatial data observed over small homogeneous spatial regions.}

{To accurately model the data's tail behavior, an alternative approach might be to rely on models justified by Extreme-Value Theory; see \citet{Davison.Huser:2015} for a general review on statistics of extremes, and \citet{Davison.etal:2012}, \citet{Davison.etal:2019} and \citet{Huser.Wadsworth:2020} for reviews on spatial extremes. Classical extreme-value models, such as max-stable processes---characterized by extreme-value copulas---and generalized Pareto processes, stem from asymptotic theory for block maxima and high threshold exceedances, respectively. However, despite their popularity, these extremal models suffer from several drawbacks: first, these limit models cannot capture weakening of dependence for increasing quantile levels. In particular, with Pareto processes, the conditional exceedance probability $\{1-2u+C(u,u)\}/(1-u)$ that appears in \eqref{eq:lambdaU} is constant in $u$ above a certain uniform quantile \citep{Rootzen.etal:2018}, while with extreme-value copulas, one has $C^k(u_1^{1/k},u_2^{1/k})=C(u_1,u_2)$ for all $(u_1,u_2)^\top\in[0,1]^2$, $k=1,2,\ldots$. While these strong restrictions on the form of the copula $C$ are indeed justified asymptotically, they may not be satisfied at finite levels (always considered in finite samples), and this has major implications in practice for assessing the risk of simultaneous extremes over a spatial region. Several recent studies have indeed shown that environmental extreme events are often found to be more spatially localized when they are more extreme \citep{Huser.Wadsworth:2019}, a property that these asymptotic extreme-value models cannot capture. Second, a consequence of these stability properties is that these extreme-value models are always tail-dependent unless they are exactly independent. As a result, non-trivial extreme-value models cannot capture independence at large distances, which makes them unsuitable over large spatial domains similarly to factor copula models. Third, these models have complicated likelihood functions that are costly to evaluate for inference \citep{Padoan.etal:2010,Castruccio.etal:2016,deFondeville.Davison:2018,Huser.etal:2019}, though recent progress on graphical models for Pareto processes opens the door to higher-dimensional likelihood inference \citep{Engelke.Hitz:2020}. Finally, because these models describe the limiting behavior of extreme events, they are typically fitted using only a small fraction of observations, thus wasting a lot of information that might potentially be useful for accurate estimation of unknown model parameters.}

{To circumvent limitations of asymptotic extreme-value models, recent work has focused on the development of ``sub-asymptotic'' models for extremes that provide additional tail flexibility at finite levels, and that can smoothly bridge both tail dependence classes under the same parametrization; see \citet{Wadsworth.Tawn:2012b}, \citet{Wadsworth.etal:2016},  \citet{Hua2017}, \citet{Su.Hua2017}, \citet{Huser.etal:2017,Huser.etal:2019,Huser.etal:2020}, \citet{Bopp.etal:2020}, and the recent review paper \citet{Huser.Wadsworth:2020}. More recently, an alternative approach based on single-site conditioning has been proposed by \citet{Wadsworth.Tawn:2019} to flexibly capture various forms of tail dependence structures, although the proposed model does not possess a convenient unconditional formulation. However, except for the rather artificial max-mixture model of \citet{Wadsworth.Tawn:2012b} and the conditional extremes model of \citet{Wadsworth.Tawn:2019}, sub-asymptotic models proposed in the literature cannot capture changes in the tail dependence class as a function of distance between sites. In other words, while it is reasonable to expect that strong tail dependence prevails at short distances and tail independence at larger distances, most proposed models in the literature assume either tail independence or tail dependence between all pairs of sites at any distance, and do not capture full independence as the distance between sites increases arbitrarily.}

%Recently, \citet{Krupskii.Huser.ea2016} and \citet{Krupskii.Genton2016} proposed some flexible and interpretable models, constructed from underlying common random factors, for non-Gaussian spatial data. However, these models can only be used to model data in a small domain as the dependence does not decrease to zero with distance. When modeling spatial data over large domains, it is important to have models that can capture tail dependence at small distances and tail independence at larger distances. Flexible models are therefore required to capture these two different types of extremal behavior at different distances.  
 
%Models that can handle different types of extremal behavior have not been extensively studied in the literature. \citet{Hua2017} and \citet{Su.Hua2017} proposed a family of bivariate copulas that allow for full range of tail dependence. \citet{Huser.Wadsworth:2019} developed a model for spatial extremes that allows for tail dependence and tail independence, and \citet{Wadsworth.Tawn2012} proposed a model for spatial data with tail independence. These models cannot, however, be used if the data are tail-dependent at smaller distances and tail-independent at larger distances. 

{In this paper, we address these shortcomings by considering process convolutions of the form
\begin{equation}
\label{eq-convmodel}
Z(\ss) = \int_{\mathbb{R}^q} k(\ss,\ss^*)W(\d \ss^*),\qquad \ss\in\mathbb{R}^q,
\end{equation} 
and variants thereof, where $k(\ss,\ss^*) \geq 0$ is a nonnegative integrable kernel function (i.e., such that $\int_{\mathbb{R}^q}k(\ss,\ss^*)\d\ss^*<\infty$ for all $\ss \in \mathbb{R}^q$) and $W$ is a particular L\'{e}vy process \citep{Sato1999} with independent increments. Note that the integral in \eqref{eq-convmodel} is a deterministic integral of Riemann-Stieltjes type, here calculated for each sample path. For simplicity, we hereafter assume that $q=2$ (unless specified otherwise), although most of our results and models can easily be extended to the case $q=1$ or $q>2$. Process convolutions have been used extensively to model spatial data; see for example \citet{Higdon2002}, \citet{Paciorek.Schervish.2006}, \citet{Calder.Cressie2007} and \citet{Zhu.Wu2010}. However, the marginal CDF of $W(\ss)$, $F_W$, is usually assumed to be Gaussian, thus leading to a Gaussian process $Z(\ss)$ in \eqref{eq-convmodel}, which does not have tail dependence. Trans-Gaussian processes $Z^*(\ss) = t\{Z(\ss)\}$, where $t(\cdot)$ is a monotone increasing transformation, could be used to model spatial data with non-Gaussian marginals \citep{Bousset.Jumel.ea2015}, but the process $Z^*(\ss)$ still possesses the Gaussian copula and thus has the same restrictive dependence structure. More flexible tail structures can be obtained by considering non-Gaussian distributions for $F_W$ in \eqref{eq-convmodel} \citep{Jonsdottir.RonnNielsen.ea2013, Noven.Veraart.ea2018}, and the unpublished manuscript of \citet{Opitz2017} investigates the dependence properties of the resulting process for an indicator kernel $k(\ss,\ss^*) = \ii\{\ss^* \in \AA(\ss)\}$ defined in terms of a hypograph indicator set $\AA(\ss)$. In this paper, we consider instead general classes of kernels in \eqref{eq-convmodel} but assume that $F_W$ is the Cauchy CDF. Because the Cauchy distribution is stable, the process $Z(\ss)$ in \eqref{eq-convmodel} remains Cauchy, which facilitates inference and theoretical calculations, and thanks to the heavy-tailedness of $F_W$, we will see that the resulting copula can have interesting tail dependence structures depending on the choice of the kernel $k$. While it would also be interesting to consider other types of (potentially skewed) stable distributions for $F_W$, we here focus on the Cauchy family, which yields tractable inference and already provides a fairly rich class of models. In this paper, we study the dependence properties of these Cauchy convolution process models under general forms of kernel functions $k$, and we derive their limiting extreme-value copulas, which turn out to be characterized in terms of a moving maximum representation. We show that a wide class of (existing or new) extremal dependence structures can be obtained, but unlike these limit extreme-value models, Cauchy convolution processes have a more flexible sub-asymptotic behavior. Moreover, when the kernel function is compactly supported, the resulting process $Z(\ss)$ has the appealing property of local tail dependence, in the sense that it possesses strong tail dependence at short distances only and full independence at larger distances. We also propose a new spatial sum-mixture model, where the proposed Cauchy process \eqref{eq-convmodel} is mixed with a lighter-tailed Gaussian process. This allows to get even higher flexibility at sub-asymptotic levels and separately control bulk and tail properties, while retaining the same extremal dependence structure. Some new theoretical results on the ``smoothness'' properties of the resulting extreme-value copulas are also derived.} 

{To make inference for the Cauchy process convolution model~\eqref{eq-convmodel} or its more flexible spatial mixture extension efficiently, we develop a fast estimation approach that consists in matching suitable empirical and model-based summary statistics. Compared to likelihood-based inference, this approach allows us to easily fit our models in higher dimensions. Unlike most extreme-value inference methods, which typically rely on extreme data from one tail only and discard all the other observations, we opt here for fitting the proposed model to the whole dataset from low to high quantiles (i.e., without applying any kind of censoring) for several reasons: first and foremost, we are not interested in modeling extremes only, but the whole distribution, as joint moderately large events from the ``bulk'' may in practice be as critical for risk assessment as individual severe extreme events; second, our approach based on the complete dataset makes full use of the available information, thus getting more accurate parameter estimates; and lastly, our proposed model is highly flexible in the bulk and the tails, so it can generally provide a good fit overall, without compromising any part of the distribution.} %For completeness, we also discuss how the corresponding extreme-value copula may be fitted by pairwise likelihood.

%The copula corresponding to the joint distribution of the process at different locations can be combined with arbitrary marginals for greater flexibility. %We show that parameters in this class of copula models can be estimated efficiently using pairwise likelihood approach.

{The rest of the paper is organized as follows. In Section \ref{sec_def}, we detail our proposed Cauchy convolution model, study its dependence properties and tail behavior, derive some interesting special cases, and discuss approximate simulation algorithms for the Cauchy convolution process itself and its extreme-value limits. We also study our proposed spatial mixture process, and explore its improved flexibility. In Section \ref{sec-inference}, we describe our proposed inference approach, while in Section \ref{sec-numericalexp}, we report the results of a simulation study, and we illustrate the proposed methodology by application to a temperature dataset from the state of Oklahoma, US. Section \ref{sec-conc} concludes with a discussion and some perspective on future research. All proofs are deferred to the Appendix.}

%%%%%%%%%%%%%%%%%%%%%%%%%%%%%%%%%%%%%%%%%%%%%%%%
%%%%%%%%%%%%%%%%%%%%%%%%%%%%%%%%%%%%%%%%%%%%%%%%
%%%%%%%%%%%%%%%%%%%%%%%%%%%%%%%%%%%%%%%%%%%%%%%%
\section{Modeling}
\label{sec_def}

%%%%%%%%%%%%%%%%%%%%%%%%%%%%%%%%%%%%%%%%%%%%%%%%
\subsection{Cauchy convolution processes and their extreme-value limits}

{We consider the process convolution \eqref{eq-convmodel}, }
{where $W$ is a L\'evy process with independent Cauchy increments, i.e., such that $W(\d\ss^*) \sim \text{Cauchy}(\d\ss^*)$ are independent increments, where Cauchy$(c)$ is the Cauchy distribution with scale parameter $c$ and probability density function (pdf) $f_{\mathcal C}(z; c) = \pi^{-1}{c/(z^2+c^2)}$, $z\in\mathbb{R}$. By some slight abuse of notation, we here write $\text{Cauchy}(\d\ss^*)$ to denote the Cauchy distribution with infinitesimal scale $c=|\d\ss^*|>0$, where $\d\ss^*$ is some infinitesimal spatial unit with area $|\d\ss^*|$.} Note that the Cauchy distribution is a sum-stable and infinitely divisible distribution, such that the process $Z(\ss)$ in \eqref{eq-convmodel} is well-defined and may be represented as
$$
Z(\ss) = \int_{\mathbb{R}^q} k(\ss,\ss^*)W^*(\ss)\d\ss^*, \quad \ss \in \mathbb{R}^q,
$$
where $W^*(\ss)$ is the Cauchy white noise process with $W^*(\ss) \sim \text{Cauchy}(1)$.

{The finite-dimensional distributions of the Cauchy process convolution $Z(\ss)$ in \eqref{eq-convmodel} are not tractable in the general case. However, it is possible to derive the extreme-value (EV) limit of this process as Proposition \ref{prop1-EVlim1} below shows. Before stating this result, we first recall some fundamentals about extreme-value theory. Let $\bm X=(X_1,\ldots,X_d)^\top$ be a $d$-dimensional random vector with margins $F_1,\ldots,F_d$ and copula $C$ as defined in the Introduction (Section~\ref{sec:intro}). The copula $C^n$ describes the copula of the vector of componentwise maxima from i.i.d.\ copies $\bm X_i=(X_{i1},\ldots,X_{id})^\top$ of $\bm X$, $i=1,\ldots,n$, i.e., $\bm M_n=(M_{n1},\ldots,M_{nd})^\top$ with $M_{nj}=\max(X_{1j},\ldots,X_{nj})$. Extreme-value copulas, denoted $C_{\rm EV}$, describe the class of dependence structures that arise as limits of $\bm M_n$ (when properly renormalized), i.e.,
\begin{equation}\label{eq:EVC}
C_{\rm EV}(u_1,\ldots,u_d)=\lim_{n\to\infty} C^n(u_1^{1/n},\ldots,u_d^{1/n}),\qquad (u_1,\ldots,u_d)^\top\in[0,1]^d.
\end{equation}
It can be shown that extreme-value copulas are such that for any $k=1,2,\ldots,$ one has $C_{\rm EV}(u_1,\ldots,u_d)=C_{\rm EV}^k(u_1^{1/k},\ldots,u_d^{1/k})$, $(u_1,\ldots,u_d)^\top\in[0,1]^d$, and they can be characterized as
\begin{equation}\label{eq:stabletail}
C_{\rm EV}(u_1,\ldots,u_d)=\exp\{-\ell(-\log u_1,\ldots,-\log u_d)\},\qquad (u_1,\ldots,u_d)^\top\in[0,1]^d,
\end{equation}
where $\ell$ is called the \emph{stable (upper) tail dependence function} and completely determines the limiting extremal dependence structure in the upper tail. From \eqref{eq:EVC} and \eqref{eq:stabletail}, the stable tail dependence function can be expressed as the limit $\ell(w_1,\ldots,w_d)=\lim_{n\to\infty}n\{1-C(1-w_1/n,\ldots,1-w_d/n)\}$, and it lies between the bounds $\max(w_1,\ldots,w_d)\leq\ell(w_1,\ldots,w_d)\leq \sum_{j=1}^d w_j$, $w_1,\ldots,w_d\geq0$, corresponding to perfect dependence and independence, respectively. Therefore, extreme-value copulas cannot be negatively dependent. As Cauchy processes are reflection-symmetric, their extremal dependence structures are identical in both tails, so hereafter we shall simply refer to $\ell$ as the \emph{stable tail dependence function}. More details on extreme-value theory, copula models, and their properties can be found, e.g., in \citet{Gudendorf.Segers:2010}, \citet{Segers:2012}, and \citet{Davison.Huser:2015}.}
{\begin{prop}[Stable tail dependence function of the Cauchy process] \label{prop1-EVlim1} \rm
Consider the Cauchy process convolution defined as in \eqref{eq-convmodel} with $W(\d\ss^*)\sim_{\text{i.i.d.}}\text{Cauchy}(\d\ss^*)$. For any collection of sites $\ss_1,  \ldots, \ss_d \in \mathbb{R}^2$, we write $Z_1 = Z(\ss_1), \ldots,$ \\$Z_d = Z(\ss_d)$. Assume that $k(\ss,\ss^*) \leq k_{\max} < \infty$ is a nonnegative bounded integrable kernel function. Let $\ell(w_1, \ldots, w_d): [0,\infty)^d \mapsto [0,\infty)$ be the stable tail dependence function of the random vector $(Z_1, \ldots, Z_d)^\top$, then
$$
	\ell(w_1, \ldots, w_d) =  \int_{\mathbb R^2} \max_{j=1,\ldots,d}w_j\zeta(\ss_j, \ss^*)\d\ss^*, \qquad  \zeta(\ss_j, \ss) = {k(\ss_j,\ss)\over\int_{\mathbb R^2} k(\ss_j,\ss^*)\d\ss^*}.  
$$
\end{prop}}

{The result of Proposition~\ref{prop1-EVlim1} implies that max-stable processes resulting from Cauchy convolution processes are from the class of moving maximum processes (see, e.g., \citet{deHaan:1984,Schlather:2002,Strokorb.etal:2015} and the references therein), whose stable tail dependence function is of form given above. Such extreme-value processes, arising as limits of properly renormalized pointwise block maxima of Cauchy convolution processes \eqref{eq-convmodel} with block size tending to infinity, admit the stochastic representation  
\begin{equation}\label{eq:movingmaximum}
Z_{\rm EV}(\ss)=\sup_{i=1,2,\ldots}\xi_i \zeta(\ss,\ss_i^*),\qquad  \zeta(\ss, \ss^*) = {k(\ss,\ss^*)\over\int_{\mathbb R^2} k(\ss,\ss^*)\d\ss^*},
\end{equation}
where $\{(\xi_i,\ss^*_i)\}$ are points from a Poisson process on $(0,\infty)\times \mathbb R^2$ with intensity $\xi^{-2}\d \xi \times \d\ss^*$. The process in \eqref{eq:movingmaximum} is max-stable and has unit Fr\'echet margins, i.e., $\Pr\{Z_{\rm EV}(\ss)\leq z\}=\exp(-1/z)$, $z>0$. A prominent example is the model introduced by \citet{Smith1990}, where the kernel has the shape of a Gaussian density, but the class is much wider than this specific example. We also note that \citet{Fasen2005} obtained a result similar to Proposition~\ref{prop1-EVlim1} for continuous-time mixed moving average processes. Similarly, \citet{Rootzen2005} studied the tail properties of stable moving average processes and established continuity of sample paths of these processes.}

{The Smith model \citep{Smith1990} is known to have very smooth sample paths; see, e.g., \citet{Davison.etal:2019}. Although this is already quite clear from the stochastic representation \eqref{eq:movingmaximum} and from spatial realizations, we now show more formally that the extreme-value limits of Cauchy processes of the form \eqref{eq-convmodel} are indeed ``smooth'' in a certain mathematical sense, and therefore that these asymptotic models may be too rigid for modeling block maxima with rough spatial dependence. ``Smoothness'' of realizations is determined by the form of dependence at short distances, and the next proposition precisely details the behavior of the stable tail dependence function for two variables from the limiting extreme-value process that are located close to each other.}
{\begin{prop}[Stable tail dependence function at short distances] \label{prop1b-EVlim1} \rm
 	Suppose that the assumptions of Proposition~\ref{prop1-EVlim1} hold. Moreover, assume that the kernel function in \eqref{eq-convmodel} may be written as $k(\ss,\ss^*) = g(\|\ss-\ss^*\|)$, where $g$ is an integrable nonnegative monotonically decreasing function. Then, for any sites $\ss_1, \ss_2\in\mathbb R^2$, the stable tail dependence function $\ell(w_1,w_2)$ of $\{Z(\ss_1),Z(\ss_2)\}^\top$ satisfies $\ell(w_1, w_2) \leq \{1+K\delta + o(\delta)\}\max(w_1, w_2)$, as $\delta:=\|\ss_1-\ss_2\| \to 0$, where $K$ is some constant that does not depend on $w_1$ and $w_2$. Furthermore, we can select $K$ such that $\ell(1,1) = 1+K\delta + o(\delta)$.
 \end{prop}}

{Let $Z_{\rm EV}(\ss)$ be the extreme-value limit \eqref{eq:movingmaximum} of the Cauchy convolution process $Z(\ss)$ defined as in \eqref{eq-convmodel} with $W(\d\ss^*)\sim_{\text{i.i.d.}}\text{Cauchy}(\d\ss^*)$, which is characterized by the stable tail dependence function $\ell$ given in Proposition~\ref{prop1-EVlim1}. To summarize the dependence structure in a spatial process, it is common to consider scale-free measures of association such as Spearman's rank correlation coefficient, $S_{\rho}(\delta)$, or the upper tail dependence coefficient defined in \eqref{eq:lambdaU}, $\lambda_U(\delta)$, here expressed as a function of the spatial distance $\delta$ between two sites. As Cauchy processes are reflection-symmetric, we have $\lambda_U(\delta)=\lambda_L(\delta)$, so we shall simply write $\lambda(\delta)$ to denote the coefficient of (lower or upper) tail dependence. Recall that while $\lambda(\delta)$ is informative about the strength of tail dependence (and is thus identical for $Z(\ss)$ and $Z_{\rm EV}(\ss)$), $S_\rho(\delta)$ mostly controls dependence in the bulk (and generally differs for $Z(\ss)$ and $Z_{\rm EV}(\ss)$). If $(Y_1,Y_2)^\top$ is a random vector distributed according to a joint distribution with continuous margins $F_1,F_2$ and underlying copula $C$, then Spearman's rank correlation is defined as ${\rm corr}\{F_1(Y_1),F_2(Y_2)\}$ and may be equivalently expressed in terms of the copula $C$ as $12\iint_{[0,1]^2} C(u_1,u_2) \d u_1 \d u_2 - 3$. The next corollary exploits Proposition \ref{prop1b-EVlim1} to show that the coefficients $\lambda(\delta)$ and $S_\rho(\delta)$ corresponding to the extreme-value copula $C_{\rm EV}$ stemming from the limiting extreme-value process $Z_{\rm EV}(\ss)$ have indeed a quite restrictive behavior at the origin, i.e., for small distances $\delta\approx0$.}
{\begin{corol}[Tail dependence and Spearman's correlation coefficients at short distances] \label{corol} \rm Under the assumptions of Proposition \ref{prop1b-EVlim1}, the limiting extreme-value process $Z_{\rm EV}(\ss)$ of the Cauchy convolution process $Z(\ss)$ (defined as in \eqref{eq-convmodel} with $W(\d\ss^*)\sim_{\text{i.i.d.}}\text{Cauchy}(\d\ss^*)$) has dependence coefficients satisfying $\lambda(\delta) = 1 - K\delta + o(\delta)$ and $S_{\rho}(\delta) > 1 - 4K\delta$, as $\delta \to 0$. 
 \end{corol}}

%\noindent
%{\emph{Proof:} It is easy to verify that $\lambda(\delta) = 2 - \ell(1,1)$, where $\ell$ is the stable tail dependence function for two sites $\ss_1,\ss_2\in\mathbb R^2$ at distance $\delta=\|\ss_1-\ss_2\|$ from each other, and by Proposition~\ref{prop1b-EVlim1}, we therefore obtain $\lambda(\delta) = 1 - K\delta + o(\delta)$. Moreover, as the function $\ell(w_1,w_2)$ in non-decreasing in both arguments and homogeneous of order $1$, we get $\ell(-\log u_1,-\log u_2) \leq -\log\{\min(u_1,u_2)\}\ell(1,1)=\log\{\min(u_1,u_2)\}\{1+K\delta+o(\delta)\}$ thanks to Proposition~\ref{prop1b-EVlim1}. This yields
%\begin{align*}
%S_{\rho}(\delta) &= 12\iint_{[0,1]^2} C_{\rm EV}(u_1,u_2) \d u_1 \d u_2 - 3 = 12\iint_{[0,1]^2} \exp\{-\ell(-\log u_1, -\log u_2)\} \d u_1 \d u_2 - 3\\
%&\geq 12\iint_{[0,1]^2} \min(u_1,u_2)^{1+K\delta}\d u_1 \d u_2 + o(\delta) - 3 = 12 {2\over 6+5K\delta+o(\delta)}-3 + o(\delta) \\
%&= {6-15K\delta+o(\delta)\over 6+5K\delta+o(\delta)}+ o(\delta) = 1 - {10K\delta\over 3} + o(\delta) > 1 - 4K\delta, \quad\delta \to 0,
%\end{align*}
%which concludes the proof. \hfill $\Box$}

{Corollary~\ref{corol} implies that moving maximum extreme-value processes resulting from Cauchy convolution processes are not suitable for modeling spatial extremes data such that $S_{\rho}(\delta)  = 1 + O(\delta^{\alpha})$, or $\lambda(\delta) = 1 + O(\delta^{\alpha})$, with $\alpha < 1$. In Section~\ref{subsec-speccase}, we describe some specific examples to illustrate this property, and in Section~\ref{subsec-convol}, we show that Cauchy convolution processes capture the sub-asymptotic dependence structure more flexibly than their extreme-value limits, and we also introduce spatial mixtures that can have different bulk and tail behaviors.}

{As already seen, the shape of the kernel $k$ in \eqref{eq-convmodel} is crucial as it determines the extremal dependence structure of Cauchy convolution processes. In the next corollary, we further show that the support of the bivariate extreme-value copula $C_{\rm EV}$ may not be the whole unit square $[0,1]^2$ depending on the kernel. This result may be used to guide the selection of a suitable kernel at a preliminary modeling stage, in order to avoid unreasonable joint behaviors.}
{\begin{corol}[The support of $C_{\rm EV}$] \label{corol2} \rm Let $\delta = ||\ss_1 - \ss_2||$ as before. Assume that the assumptions of Proposition \ref{prop1b-EVlim1} hold, and that 
$$G(\delta) = \max_{(s_1^*,s_2^*)^\top\in\mathcal{S}_{\cup+}(\delta)} g(\|(s_1^* + \delta, s_2^*)^\top\|)/g(\|(s_1^*, s_2^*)^\top\|)  < \infty,$$
where $\mathcal{S}_{\cup+}(\delta):=\{(s_1^*,s_2^*)\in\mathbb R^2:g(\|(s_1^*, s_2^*)^\top\|)>0\text{ and/or } g(\|(s_1^* + \delta, s_2^*)^\top\|)>0\}$, with the convention that $x/0=\infty$ for all $x>0$. Then, $\ell(w_1, w_2) = w_1$ for all $(w_1,w_2)^\top\in[0,\infty)^2$ with $w_1/w_2 > G(\delta)\geq1$, and similarly,  $\ell(w_1, w_2) = w_2$ for all $(w_1,w_2)^\top\in[0,\infty)^2$ with $w_2/w_1 > G(\delta)\geq1$, which implies that the extreme-value copula $C_{\rm EV}(u_1,u_2)$ has density zero on the region defined by $\{(u_1,u_2)^\top\in[0,1]^2:u_1 < u_2^{G(\delta)}\;\;\text{or}\;\;u_2 < u_1^{G(\delta)}\}$.
\end{corol}
}

%\noindent
%{\emph{Proof:} If $w_1/w_2 > G(\delta)$, then $S(w_1, w_2) = \{(s_1^*,s_2^*)^\top\in\mathbb R^2: w_2 g(\|(s_1^*+\delta,s_2^*)^\top\|) > w_1g(\|(s_1^*,s_2^*)^\top\|)\}$ is the empty set, i.e., $S(w_1,w_2)=\emptyset$, which implies that $\ell(w_1,w_2) = w_1$ by following the proof of Proposition~\ref{prop1b-EVlim1}. The second part of the corollary follows by symmetry. \hfill $\Box$}

{In particular, for $g(t) = \exp(-t^\alpha)$, $t\geq0$, with $0<\alpha \leq 1$,  we obtain by the triangle inequality that
\begin{align*}
G(\delta) &\leq \max_{(s_1^*,s_2^*)\in\mathbb R^2} \exp\{\|(s_1^*, s_2^*)^\top\|^{\alpha} - \|(s_1^*+\delta, s_2^*)^\top\|^{\alpha} \}\\
&\leq \exp\{\|(s_1^*, s_2^*)^\top\|^{\alpha} - \|(s_1^*, s_2^*)^\top\|^{\alpha} + \delta^{\alpha} \} = \exp(\delta^{\alpha}),
\end{align*}
which is finite for all $\delta\geq0$. Thus, Corollary~\ref{corol2} implies that for all kernels of the form $k(\ss,\ss^*)=\exp\{-(\|\ss-\ss^*\|/\lambda)^\alpha\}$, with $\lambda>0$ and $0<\alpha\leq1$, the extreme-value copula $C_{\rm EV}$ resulting from Cauchy convolution processes does not have full support, thus preventing ``very low extreme values'' at one location to occur with ``very high extreme values'' at another location. 
By contrast, it is easy to verify that $G(\delta)=\infty$ for all $\alpha > 1$, and for all kernels $k(\ss,\ss^*)=g(\|\ss-\ss^*\|)$ that are compactly supported, i.e., such that $g(t)=0$ for all $t>r$ for some range $r>0$. This odd behavior is illustrated in Section~\ref{subsec-simul}. In practice, it may be sensible to restrict ourselves to $\alpha>1$ when using  $k(\ss,\ss^*)=\exp\{-(\|\ss-\ss^*\|/r)^\alpha\}$, or to use a different (potentially compactly supported) kernel, to avoid pathological behaviors.}
%%%%%%%%%%%%%%%%%%%%%%%%%%%%%%%%%%%%%%%%%%%%%%%%
\subsection{Special cases}
\label{subsec-speccase}
{We now give some specific examples that have a tractable bivariate extreme-value copula $C_{\rm EV}$.}
{\begin{example}[Marshall--Olkin copula]
Consider the indicator kernel $k(\ss,\ss^*) = \ii\{\ss^* \in \AA(\ss)\}$, where $\AA(\ss) \subset \mathbb{R}^q$ is a compact subset of $\mathbb R^q$ with area $0 < |\AA(\ss)| < \infty$ for each $\ss \in \mathbb{R}^q$. From Proposition~\ref{prop1-EVlim1}, it is easy to check that the stable tail dependence function of $\{Z(\ss_1),Z(\ss_2)\}^\top$ is
%\begin{align*}
%\ell(w_1,w_2) =& \,{|\AA(\ss_1)\backslash \AA(\ss_2)|\over |\AA(\ss_1)|}\cdot w_1 + {|\AA(\ss_2)\backslash \AA(\ss_1)|\over |\AA(\ss_2)|}\cdot w_2 \\
%&+ |\AA(\ss_1)\cap \AA(\ss_2)|\cdot \max \left\{{w_1\over |\AA(\ss_1)|}, {w_2\over|\AA(\ss_2)|}  \right\}.
%\end{align*}
\begin{multline*}
\ell(w_1,w_2) = {|\AA_1\backslash \AA_2|\over |\AA_1|}\cdot w_1 + {|\AA_2\backslash \AA_1|\over |\AA_2|}\cdot w_2 + |\AA_1\cap \AA_2|\cdot \max \left\{{w_1\over |\AA_1|}, {w_2\over|\AA_2|}  \right\},
\end{multline*}
where we have written $\AA_i=\AA(\ss_i)$, $i=1,2$, for simplicity. 
The corresponding limiting extreme-value process is thus driven by the Marshall--Olkin copula with a singular component \citep{Marshall.Olkin1967}.
\end{example}}

{It is also possible to obtain the extreme-value limit of the Cauchy convolution process in closed form for non-trivial kernels in some special cases. In particular, one can consider the stationary kernel $k(\ss,\ss^*) = g(\|\ss-\ss^*\|)$ where $g$ is a nonnegative continuous function.}

\begin{example}[Smith model \citep{Smith1990}]
{Assume that $q=1$ and that $k(s,s^*) = \phi(s-s^*; \sigma^2)$, where $\phi(\cdot; \sigma^2)$ is the Gaussian density function with mean zero and variance $\sigma^2$. For simplicity, let also assume that $s_1 < s_2$. Adopting the notation of Proposition~\ref{prop1b-EVlim1}, it follows that $\{s^*\in\mathbb R: w_1\zeta(s_1, s^*) > w_2\zeta(s_2, s^*)\} = \{s^*\in\mathbb R:  s^* <  {\sigma^2\over s_2 - s_1}\log(w_1/w_2) + {s_1+s_2\over2}\}$, and thus
$$
\ell(w_1,w_2) = w_1\Phi\left({\lambda^*_1\over 2} + {1\over \lambda^*_1}\log{w_1\over w_2}\right) + w_2\Phi\left({\lambda^*_1\over 2} + {1\over \lambda^*_1}\log{w_2\over w_1}\right), \quad \lambda^*_1 = {|s_1-s_2|\over\sigma},
$$
where $\Phi(\cdot)$ denotes the standard normal CDF. The limiting extreme-value copula is thus the H\"{u}sler--Reiss copula \citep{Husler.Reiss1989}. Here, we can easily verify that $\lambda(\delta)=2-\ell(1,1)=1+O(\delta)$, as $\delta=|s_1-s_2|\to0$. Note that $\lambda_1^* = \{\gamma(s_1-s_2)\}^{1/2}$, where $\gamma(h) = (|h|/\sigma)^2$ is a valid variogram. This model corresponds to the Smith max-stable model \citep{Smith1990}, which is a smooth limiting case of the Brown--Resnick model \citep{Kabluchko.etal:2009} with variogram $\gamma(h)=(|h|/\sigma)^\alpha$, $\sigma>0,\alpha\in(0,2)$.} More details for the case of $q=2$ are provided in the Appendix.

\end{example}
{\begin{example}[Laplace kernel in $\mathbb{R}$]
Assume that $q=1$ and let $k(s,s^*) = $\\ ${1\over 2\lambda}\exp\left(-{|s-s^*|/\lambda}\right)$. For simplicity assume $s_1 < s_2$. It follows that for $G_\lambda(\delta) = \exp(\delta/\lambda)$ with $\delta=|s_1-s_2|$ and $M(s_1, s_2) = \{s^*\in\mathbb R: w_1\zeta(s_1, s^*) \geq w_2\zeta(s_2, s^*)\}$, 
$$M(s_1, s_2) = \begin{cases}
\left\{s^*\in\mathbb R: s^* \leq {\lambda\over2}\log({w_1/w_2}) + {s_1+s_2\over 2}\right\}, & {1\over G_\lambda(\delta)} \leq {w_1\over w_2} \leq G_\lambda(\delta),\\
\mathbb{R}, & {w_1\over w_2} > G_\lambda(\delta),\\
\emptyset, & {w_1\over w_2} <{1\over G_\lambda(\delta)},
\end{cases}$$
and, for any sites $s_1, s_2\in\mathbb R$, 
$$
\ell(w_1, w_2) = \begin{cases}
w_1+w_2 - \sqrt{{w_1w_2/G_\lambda(\delta)}},& {1\over G_\lambda(\delta)} \leq {w_1\over w_2} \leq G_\lambda(\delta),\\
w_1,& {w_1\over w_2} > G_\lambda(\delta),\\
w_2,& {w_1\over w_2} < {1\over G_\lambda(\delta)}. 
\end{cases}
$$
Thus, the corresponding extreme-value copula density is positive only when \\$u_2^{G_\lambda(\delta)} < u _1 < u_2^{1/G_\lambda(\delta)}$. For this model, we can again easily verify that $\lambda(\delta)=1+O(\delta)$ as $\delta=|s_1-s_2 |\to0$.
\end{example}}

\begin{example}[Kernels with compact support]
{Notice that if the set $\mathcal S_{\cap+}(\ss_1,\ss_2) :=\{\ss^*\in\mathbb R^q: k(\ss_1,\ss^*) > 0, k(\ss_2,\ss^*) > 0\} = \emptyset$ for some sites $\ss_1, \ss_2\in\mathbb R^q$, then $\ell(w_1,w_2) = w_1 + w_2$. In particular, if the kernel $k$ is compactly supported such that $k(\ss,\ss^*) = 0$ whenever $\|\ss- \ss^*\| > r$ for some radius $r>0$, then $\mathcal S_{\cap+}(\ss_1, \ss_2)$ is empty if and only if $\|\ss_1 - \ss_2\| \geq 2r$. This implies that the variables $Z_{\rm EV}(\ss_1)$ and $Z_{\rm EV}(\ss_2)$ from the limiting extreme-value process are independent whenever $\|\ss_1 - \ss_2\| \geq 2r$. By construction, two realizations $Z(\ss_1)$ and $Z(\ss_2)$ with $\|\ss_1 - \ss_2\| \geq 2r$ from the process convolution in \eqref{eq-convmodel} are not only tail-independent but fully independent in this case. A flexible family of compactly supported kernels includes $k(\ss,\ss^*)=\{1-(\|\ss-\ss^*\|/r)^\alpha\}_+^\eta$, where $a_+=\max(0,a)$, for some parameters $r,\alpha,\eta>0$, though for identifiability concerns one may fix either $\alpha$ and/or $\eta$ in practice. Here, the parameter $r$ defines the range of spatial dependence for this process and should not be fixed.} Another example with compactly supported kernel is detailed in the Appendix.

\end{example}

%%%%%%%%%%%%%%%%%%%%%%%%%%%%%%%%%%%%%%%%%%%%%%%%
\subsection{Mixture of Cauchy and Gaussian processes}
\label{subsec-convol}

{The Cauchy convolution process $Z(\ss)$ defined as in \eqref{eq-convmodel} with $W(\d\ss^*)\sim_{\text{i.i.d.}}\text{Cauchy}(\d\ss^*)$ has the appealing property of being tail-dependent (unless exactly independent) and the strength of dependence as a function of distance between two spatial locations is controlled by the kernel function $k(\ss,\ss^*)$. In particular, if the kernel $k$ has a compact support, the process is only dependent locally (i.e., at small distances), and is independent at large enough distances. Moreover, importantly, given that the Cauchy convolution process $Z(\ss)$ can be seen as a sum-mixture of heavy-tailed noise, rather than a max-mixture like its extreme-value limit $Z_{\rm EV}(\ss)$ (recall Proposition~\ref{prop1-EVlim1} and \eqref{eq:movingmaximum}) or the max-mixture models proposed by \citet{Wadsworth.Tawn:2012b}, it can generate more flexible patterns and realistic realizations, as demonstrated below. However, Cauchy convolution processes also have certain drawbacks when modeling spatial data, namely: 
\begin{enumerate}
\item Like Gaussian processes, the resulting copula is reflection-symmetric (and in particular, tail-symmetric), which might not be realistic in some applications;
\item Depending on the kernel $k$ and the distance between sites $\ss_1,\ss_2$, two variables $Z(\ss_1)$ and $Z(\ss_2)$ can either be tail-dependent, or exactly independent, but the intermediate case of (non-trivial) tail independence is not possible. In other words, the process cannot capture tail independence (unless exactly independent) and thus, it still lacks flexibility at sub-asymptotic levels;
\item The strength of dependence in the bulk of the joint distribution of \\$\{Z(\ss_1), \ldots, Z(\ss_d)\}^\top$ and in its tails cannot be controlled separately with this process.
\end{enumerate}}

\begin{figure}[t!]
	\begin{center}
		\includegraphics[width=0.32\linewidth]{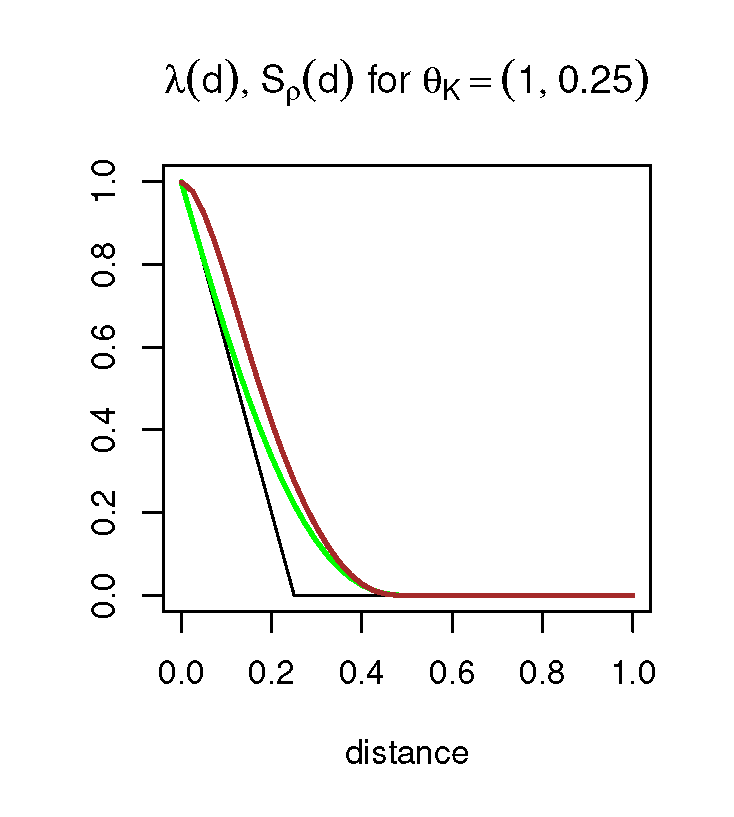}
		\includegraphics[width=0.32\linewidth]{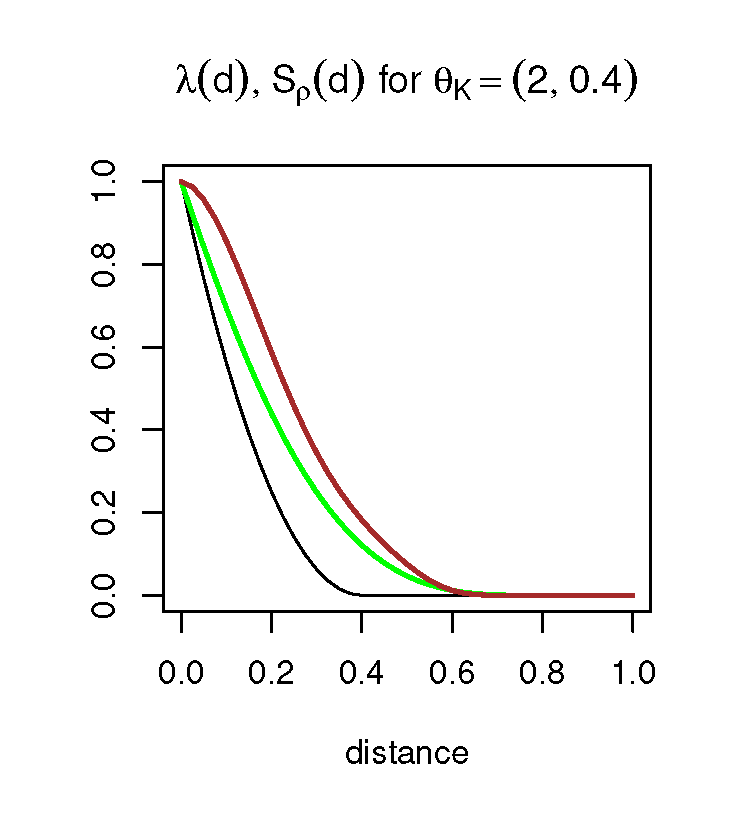}
		\includegraphics[width=0.32\linewidth]{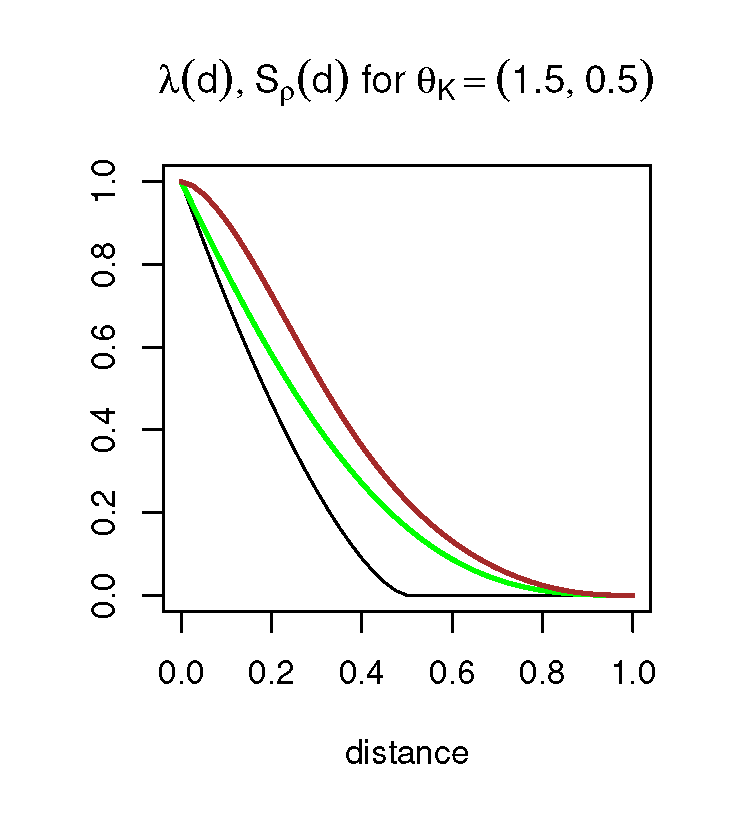}
		\caption{{\footnotesize {Coefficient of upper tail dependence, $\lambda(\delta)$ (green line), and Spearman's rank correlation coefficient, $S_{\rho}(\delta)$ (brown line), for the Cauchy convolution process $Z(\ss)$ defined as in \eqref{eq-convmodel} with kernel function $k(\ss,\ss^*;\tht_K) = (1-\|\ss-\ss^*\|/r)_+^{\eta}$ (black line) and kernel parameters $\tht_K =(\eta, r)^\top = (1, 0.25)^\top$ (left), $\tht_K = (2, 0.4)^\top$ (middle) and $\tht_K = (1.5, 0.5)^\top$ (right).}}}
		\label{fig1notflex}
	\end{center}
\end{figure}

While the issue highlighted in the first point is application-specific and should be addressed in future research, we have not found it to be a major limitation in our temperature data application described in Section~\ref{sec-empstudy}. The second and third points highlight, however, issues that are more critical from a risk assessment perspective where the tail dependence structure needs to be estimated with accuracy. 
In applications, it is common to observe very weak or zero tail dependence at large distances, while fairly strong overall dependence prevails in the bulk. In other words, it can happen in practice that the data suggest both $\lambda(\delta) \approx 0$ and $S_{\rho}(\delta) \gg 0$ for reasonably large distances $\delta$, but the Cauchy convolution process \eqref{eq-convmodel} cannot capture this situation, as the strength and range of dependence in the bulk and the tails are necessarily similar to each other, and cannot be controlled separately, as illustrated in Figure~\ref{fig1notflex}.%%% and further discussed in the Supplementary Material.}

{To increase flexibility of the Cauchy process convolution model and circumvent the issues highlighted in the second and third points above, we propose to modify the original process by mixing it with a tail-independent process possessing lighter tails. Specifically, we define
\begin{equation}
\label{eq-convgauss}
\tilde{Z}(\ss) = Z(\ss) + \beta Z_G(\ss),
\end{equation}
where $Z(\ss)$ is the Cauchy process \eqref{eq-convmodel} with $W(\d\ss^*)\sim_{\text{i.i.d.}}\text{Cauchy}(\d\ss^*)$, $Z_G(\ss)$ is a stationary Gaussian process with standard normal $N(0,1)$ marginals and some correlation function $\rho_G(\delta)$, $\delta\geq0$, and $\beta\geq0$. For simplicity, we assume that $\rho_G(\delta)$ is an isotropic correlation function, though all the results in this section can be readily extended to the anisotropic or non-stationary context. As the next proposition shows, the new process possesses the same asymptotic behavior as the process (\ref{eq-convmodel}), such that the spatial mixture model (\ref{eq-convgauss}) still captures local tail dependence if the kernel is compactly supported.}
{\begin{prop}[Tail behavior of the spatial mixture process $\tilde Z(\ss)$] \label{prop2-EVlim1} \rm
	Under the assumptions of Proposition \ref{prop1-EVlim1}, the stable tail dependence function corresponding to the joint distribution of $\{\tilde{Z}(\ss_1), \ldots, \tilde{Z}(\ss_d)\}^\top$, with the process $\tilde{Z}(\ss)$ defined as in \eqref{eq-convgauss}, is $\ell(w_1, \ldots, w_d)$ given in Proposition~\ref{prop1-EVlim1}.  	
\end{prop}}

%%%PK: remove this part to save space
{While non-trivial tail independence can be captured by the spatial mixture process, it is important to remark that the tail dependence structure still remains fairly restricted in this case. To see this, assume that the random variables $Z(\ss_1), \ldots, Z(\ss_d)$ from the Cauchy process appearing in \eqref{eq-convgauss} are independent. Then, from the proof of Proposition \ref{prop2-EVlim1}, we get that
	$$p_{L,n} + o(e^{-n/(2\beta^2)}) \leq \Pr(\tilde Z_1 \leq n, \ldots, \tilde Z_d \leq n) \leq p_{U,n} + o(e^{-n/(2\beta^2)}),
	$$
where 
$p_{L,n}=\Pr(Z_1 \leq n - \sqrt{n}, \ldots, Z_d \leq n - \sqrt{n}) = \{\Pr(Z_1 \leq n)\}^d + o(\{\Pr(Z_1 \leq n)\}^d)=\{\Pr(\tilde Z_1 \leq n)\}^d + o(\{\Pr(\tilde Z_1 \leq n)\}^d)$, and $p_{U,n}=\Pr(Z_1 \leq n + \sqrt{n}, \ldots, Z_d \leq n + \sqrt{n}) = \{\Pr(Z_1 \leq n)\}^d + o(\{\Pr(Z_1 \leq n)\}^d)=\{\Pr(\tilde Z_1 \leq n)\}^d + o(\{\Pr(\tilde Z_1 \leq n)\}^d)$, 
and this implies that the tail order of a copula linking $\tilde Z(\ss_1), \ldots, \tilde Z(\ss_d)$ is equal to $\kappa = d$. In other words, this copula has tail quadrant independence, which is a weak form of tail independence. Nevertheless, the spatial mixture process \eqref{eq-convgauss} enjoys great flexibility at sub-asymptotic levels, as demonstrated below.}

\begin{figure}[t!]
	\begin{center}
		\includegraphics[width=0.9\linewidth]{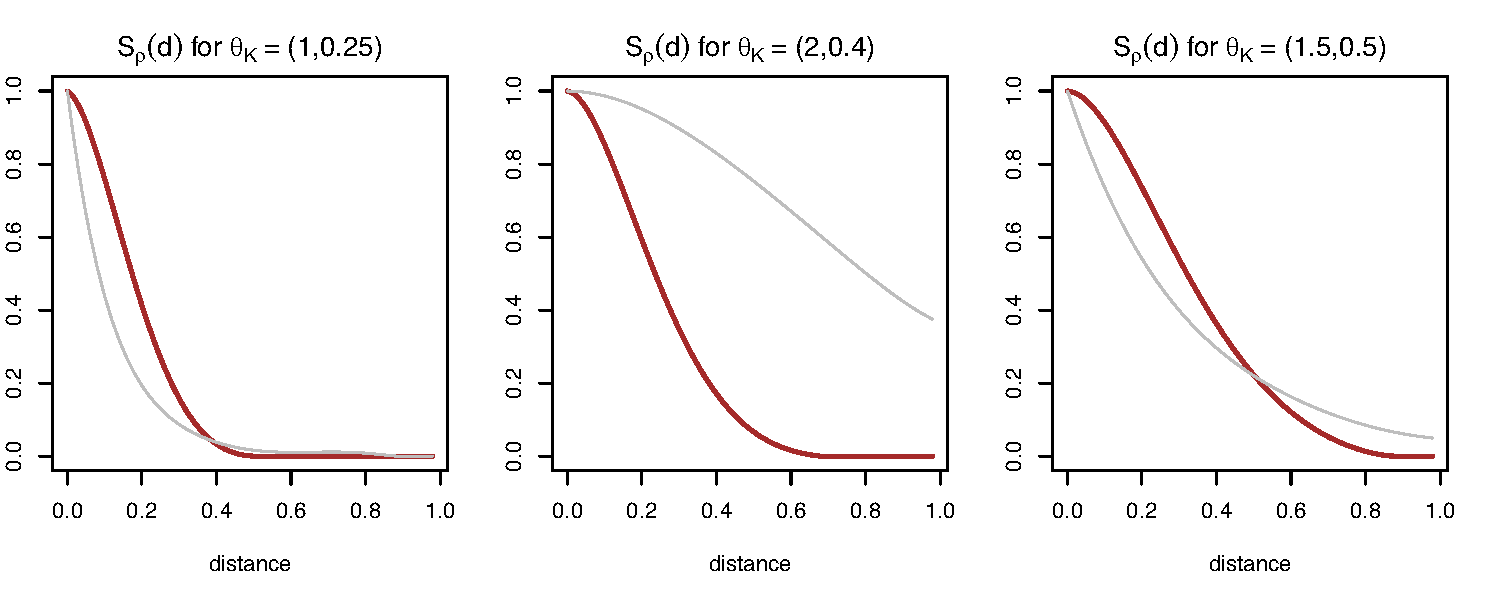}
		\caption{{\footnotesize {Spearman's correlation coefficient, $S_{\rho}(\delta)$, for the Cauchy convolution process $Z(\ss)$ defined in \eqref{eq-convmodel} with $W(\d\ss^*)\sim_{\text{i.i.d.}}\text{Cauchy}(\d\ss^*)$ (brown), and for the process $\tilde{Z}(\ss)$ defined in \eqref{eq-convgauss} (grey), based on the kernel function $k(\ss,\ss^*;\tht_K) = (1-\|\ss-\ss^*\|/r)_+^{\eta}$ with $\tht_K =(\eta, r)^\top = (1, 0.25)^\top$ and $\rho_G(\delta) = \exp(-8 d)$ (left), $\tht_K  = (2, 0.4)^\top$ and $\rho_G(\delta) = \exp(- \delta^2)$ (middle) and $\tht_K =(1.5, 0.5)^\top$ and $\rho_G(\delta) = \exp(-3 \delta)$ (right). For the model $\tilde{Z}(\ss)$, we set $\beta=2$. }}}
		\label{fig1flex}
	\end{center}
\end{figure}

To illustrate the improved flexibility of model (\ref{eq-convgauss}), Figure~\ref{fig1flex} shows $S_{\rho}(\delta)$ for the process $\tilde{Z}(\ss)$ in \eqref{eq-convgauss} with various kernel functions, $\beta=2$, and underlying correlation function for the Gaussian process equal to $\rho_G(\delta) = \exp(-8 \delta)$, $\exp(-\delta^2)$, or $\exp(-3 \delta)$, respectively.   
We can see that $\rho_G(\delta)$ mainly controls the Spearman's correlation of the process $\tilde{Z}(\ss)$, whereas the kernel parameters $\tht_K$ control its tail dependence structure, as expected. The new spatial mixture process therefore allows for a greater flexibility both in the tails and in the middle of the joint distribution, capturing a wide range of sub-asymptotic dependence behaviors. %%; see the Supplementary Material for further comments.} %Various types of correlation functions $\rho_G(\delta)$ can be used to further increase the flexibility of the model \eqref{eq-convgauss}. For example, tapered covariance functions can be used to control the range of the overall dependence and kernels with finite support can be selected to control the range of tail dependence of the spatial process. Correlation functions $\rho_G(\delta)$ with a smoothness parameter can be used to control the smoothness of realizations of the process $\tilde{Z}(\ss)$ in \eqref{eq-convgauss} (see Figure~\ref{fig1flex}, left to right).}

%%%%%%%%%%%%%%%%%%%%%%%%%%%%%%%%%%%%%%%%%%%%%%%%
\subsection{Simulation and realizations}
\label{subsec-simul}
{Fast approximate simulation from the Cauchy convolution process $Z(\ss)$ defined in \eqref{eq-convmodel} with $W(\d\ss^*)\sim_{\text{i.i.d.}}\text{Cauchy}(\d\ss^*)$ can be easily obtained at locations $\ss_1, \ldots, \ss_d\in\mathbb R^2$ as $Z(\ss_j) \approx \delta_m^2 \sum_{k,l=1}^m k(\ss_j, \ss_{k,l}) W_{k,l}$, $W_{k,l} \sim_{\text{i.i.d}} \text{\text{Cauchy}}(1)$, $j=1,\ldots,d$, where $\{\ss_{k,l}\}_{k,l=1}^m$ is a fine rectangular grid in $S_m \subset \mathbb R^2$, with the cell size $\delta_m \times \delta_m$, covering the simulation sites, with large $m$ (see the proof of Proposition~\ref{prop1-EVlim1}). Similarly, the process $\tilde{Z}(\ss)$ in \eqref{eq-convgauss} can be simulated based on a finite approximation as $\tilde{Z}(\ss_j) \approx \delta_m^2 \sum_{k,l=1}^m k(\ss_j, \ss_{k,l}) W_{k,l} + \beta Z_G(\ss_j)$, $W_{k,l} \sim_{\text{i.i.d}} \text{\text{Cauchy}}(1)$, $j=1,\ldots,d$, where $\{Z_{G}(\ss_1),\ldots,Z_{G}(\ss_d)\}^\top$ is a realization from the Gaussian random field $Z_G(\ss)$ at the sites $\ss_1,\ldots,\ss_d$, independent of the Cauchy variables $\{W_{k,l}\}$.}

{Simulation from the associated limiting extreme-value process $Z_{\rm EV}(\ss)$ is also quite straightforward. Given that the kernel function $k(\ss,\ss^*)$ is bounded, simulating moving maximum max-stable random fields can be performed exactly by exploiting the stochastic representation~\eqref{eq:movingmaximum}%, simulating the Poisson points $\{\xi_i\}$ in decreasing order and setting an appropriate stopping rule
; see \citet{Schlather:2002}. If the kernel is compactly supported with range $r>0$, then the study region should be expanded on all sides by $r$ units, and if it has support over $\mathbb R^2$, then the study region should be expanded sufficiently to ensure that the contributions of ``distant'' points in \eqref{eq:movingmaximum} become negligible. Alternatively, fast approximate simulations are also possible; see the Appendix for more details.}

Figure~\ref{fig1fct} shows realizations of the processes $Z(\ss)$, $\tilde{Z}(\ss)$ and $Z_{\rm EV}(\ss)$ in $[0,1]^2$ obtained for different kernel functions. %using Proposition~\ref{prop1-EVlim2}. 
For the indicator kernel with compact support we used a rectangular grid on $[-r, 1+r]^2$ to prevent edge effects. For kernels with infinite support, we used a regular grid on $S_m = [-1, 2]^2$ because $k(\ss,\ss^*) < 5\cdot10^{-5}$ for $\|\ss-\ss^*\| > 1$ for the considered kernels. We use %the approximate algorithm from Proposition~\ref{prop1-EVlim2} with 
$m=200$ in all cases. 
\begin{figure}[t!]
	\begin{center}
		\includegraphics[width=\linewidth]{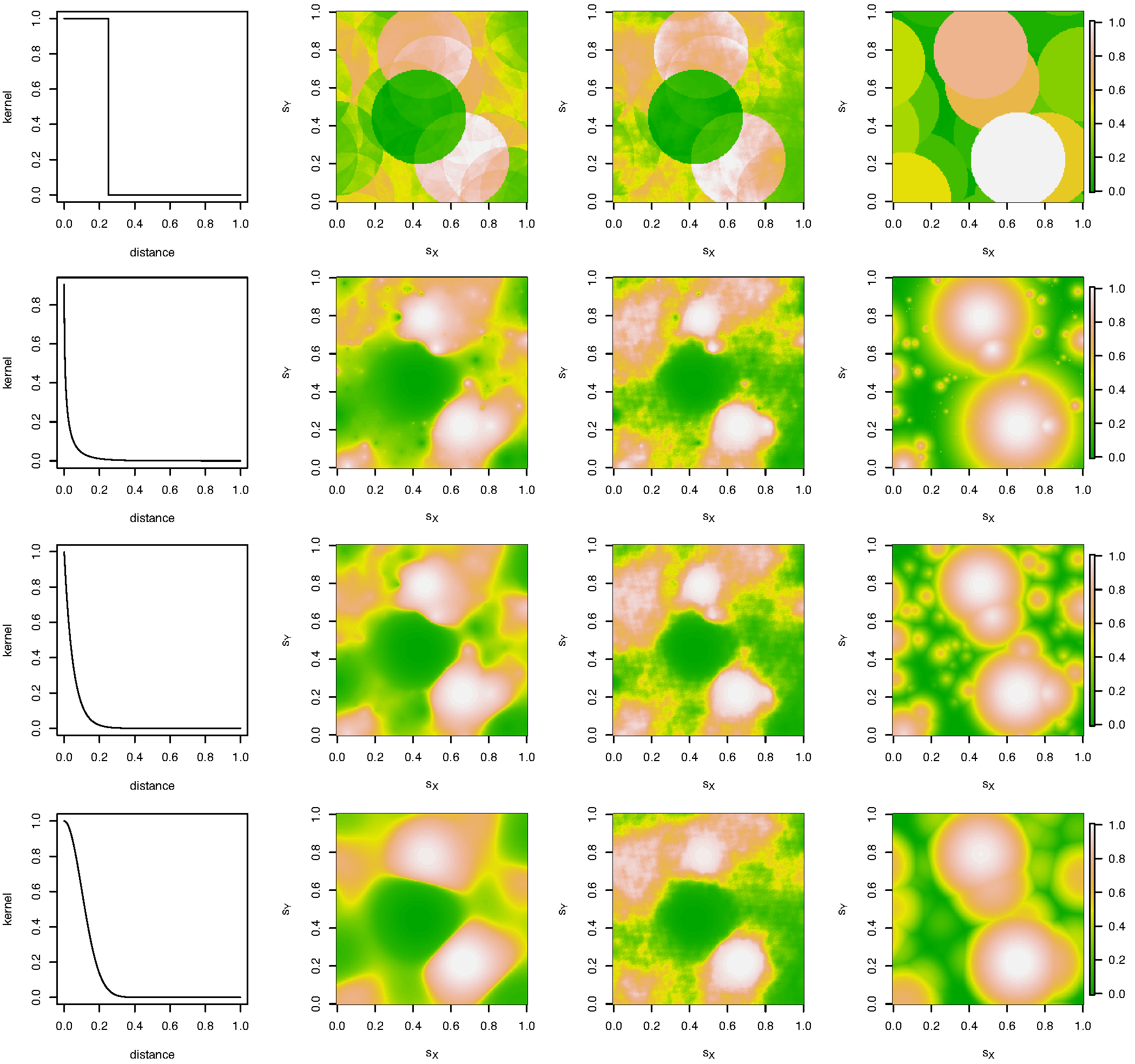}\vspace{-5mm}
		\caption{{\footnotesize {Realizations on $[0,1]^2$ from the Cauchy convolution process $Z(\ss)$ defined in \eqref{eq-convmodel} with $W(\d\ss^*)\sim_{\text{i.i.d.}}\text{Cauchy}(\d\ss^*)$ ($2^{\rm nd}$ column), $\tilde{Z}(\ss)$ defined in \eqref{eq-convgauss} ($3^{\rm rd}$ column), and their extreme-value limit $Z_{\rm EV}(\ss)$ ($4^{\rm th}$ column) for different isotropic kernels $k(\ss,\ss^*)=g(\|\ss-\ss^*\|)$ (the $1^{\rm st}$ column shows the function $g$). Each simulated process is marginally transformed to have ${\rm Unif}(0,1)$ marginals. The same random seed was used for all realizations. The kernel is $k(\ss,\ss^*) = \ii(\|\ss-\ss^*\| < 0.25)$ ($1^{\rm st}$ row), $k(\ss,\ss^*) = \exp\left(-10\|\ss-\ss^*\|^{1/2}\right)$ ($2^{\rm nd}$ row), $k(\ss,\ss^*) = \exp\left(-20\|\ss-\ss^*\|\right)$ ($3^{\rm rd}$ row), and $k(\ss,\ss^*) = \exp\left(-50\|\ss-\ss^*\|^{2}\right)$ ($4^{\rm th}$ row). We use $\beta=2$ and $\rho_G(\delta) = \exp(-\delta)$ for $\tilde{Z}(\ss)$.}}}
		\label{fig1fct}
	\end{center}
\end{figure}
%%%\vspace{-.82cm}
We can see that the realizations from the Cauchy process $Z(\ss)$ (second column) can indeed generate quite complex patterns, as opposed to its fairly rigid extreme-value limit $Z_{\rm EV}(\ss)$ (fourth column). Furthermore, the spatial mixture process $\tilde{Z}(\ss)$ (third column) has much rougher spatial realizations than the Cauchy process $Z(\ss)$, confirming the higher degree of flexibility of $\tilde{Z}(\ss)$ to capture the sub-asymptotic dependence structure. Moreover, in all simulations, the effect of the kernel on the spatial dependence structure is obvious, especially with the extreme-value process $Z_{\rm EV}(\ss)$ (fourth column). Clearly, smoother kernels result in smoother random fields. 

\begin{figure}[t!]
	\begin{center}
		%%2.0, 2.3, 2.7, 2.3
		%\includegraphics[width=0.24\linewidth]{fig_sp_indic.png}\hspace{-3mm}
		%\includegraphics[width=0.24\linewidth]{fig_sp_powexp2.png}\hspace{-3mm}%indicator kernel
		%\includegraphics[width=0.24\linewidth]{fig_sp_exp.png}\hspace{-3mm}
		%\includegraphics[width=0.24\linewidth]{fig_sp_gauss.png}
		\includegraphics[width=\linewidth]{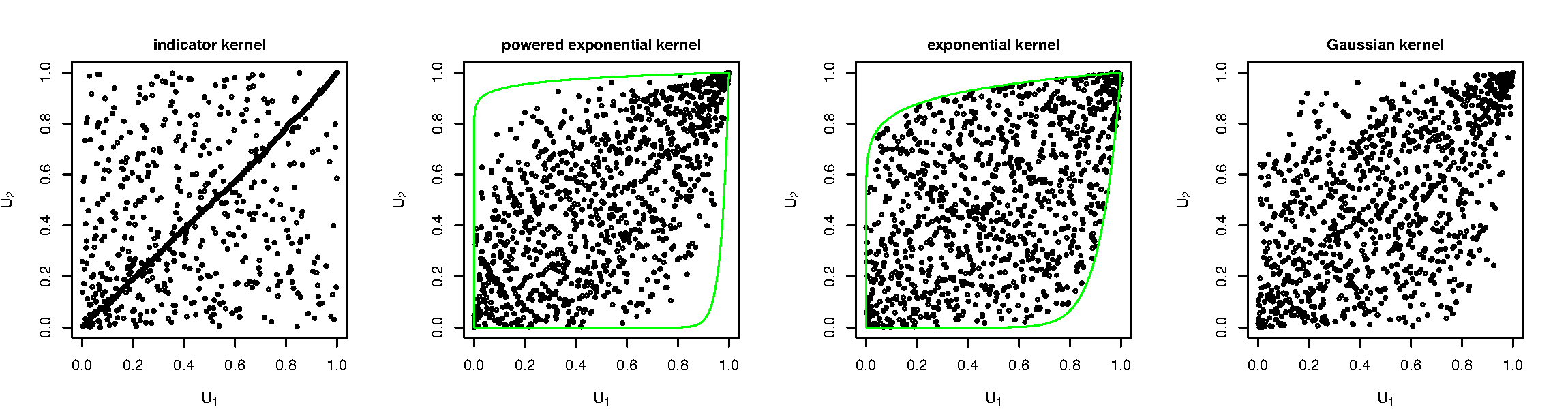}
		\caption{{\footnotesize {Bivariate scatter plots of datasets of size $1000$ generated from the extreme-value copulas linking $Z_{\rm EV}(\ss_1)$ and $Z_{\rm EV}(\ss_2)$ at distance $\delta=\|\ss_1 - \ss_2\| = 0.125$ based on the kernel $k(\ss,\ss^*) = \ii(\|\ss-\ss^*\| < 0.25)$, $k(\ss,\ss^*) = \exp\left(-10\|\ss-\ss^*\|^{1/2}\right)$, $k(\ss,\ss^*) = \exp\left(-20\|\ss-\ss^*\|\right)$, and $k(\ss,\ss^*) = \exp\left(-50\|\ss-\ss^*\|^{2}\right)$ (left to right). Green lines $u_1 = u_2^{\pm\exp(10\delta^{1/2})}$ and $u_1 = u_2^{\pm\exp(20\delta)}$ delimit the area with positive density in the $2^{\rm nd}$ and $3^{\rm rd}$ panels, respectively.}}}
\label{fig1sp}
\end{center}
\end{figure}

Figure~\ref{fig1sp} shows bivariate scatter plots for datasets of size $1000$ generated from the extreme-value copulas linking $Z_{\rm EV}(\ss_1)$ and $Z_{\rm EV}(\ss_2)$ at distance $\|\ss_1 - \ss_2\| = 0.125$ for the four extremal processes displayed in Figure~\ref{fig1fct}.
With the indicator kernel, the singular component on the diagonal $u_1 = u_2$ is clearly seen. Green lines $u_1 = u_2^{\pm\exp(20\delta)}$ and $u_1 = u_2^{\pm\exp(10\delta^{0.5})}$, with $\delta=0.125$, delimit the area with positive density for the data generated using the exponential and powered exponential kernels; recall Corollary~\ref{corol2}. Note that for the powered exponential kernel, the copula density is positive but very small near the boundaries. By contrast, the Gaussian kernel yields a positive density on the whole unit square $[0,1]^2$, and the indicator kernel has a positive density on $[0,1]^2\setminus\{(u_1,u_2)^\top:u_1=u_2\}$.

In the next section, we detail our proposed inference approach for the Cauchy convolution process $Z(\ss)$ observed at $d$ locations $\ss_1, \ldots, \ss_d$, and for the modified spatial mixture process $\tilde{Z}(\ss)$.%, and for their corresponding extreme-value limit, $Z_{\rm EV}(\ss)$.}

%%%%%%%%%%%%%%%%%%%%%%%%%%%%%%%%%%%%%%%%%%%%%%%%
%%%%%%%%%%%%%%%%%%%%%%%%%%%%%%%%%%%%%%%%%%%%%%%%
%%%%%%%%%%%%%%%%%%%%%%%%%%%%%%%%%%%%%%%%%%%%%%%%
\section{Inference}
\label{sec-inference}

%%%%%%%%%%%%%%%%%%%%%%%%%%%%%%%%%%%%%%%%%%%%%%%%
\subsection{General setting and marginal estimation}
\label{subsec-est-0}
{Throughout this section, we assume that $\{\yy_i = (y_{i1},\ldots,y_{id})^\top\}_{i=1}^n$ are $n$ i.i.d.\ realizations of a stationary process $Y(\ss)$ measured at $d$ spatial locations $\ss_1, \ldots, \ss_d\in\mathbb R^2$, and whose dependence structure (i.e., copula) is the same as either (i) the Cauchy convolution process $Z(\ss)$ defined as in \eqref{eq-convmodel} with $W(\d\ss^*)\sim_{\text{i.i.d.}}\text{Cauchy}(\d\ss^*)$ using some parametric kernel function $k$ (Section~\ref{subsec-est-1}); or (ii) the spatial mixture model extension $\tilde{Z}(\ss)$ defined as in \eqref{eq-convgauss} (Section~\ref{subsec-est-3}); or (iii) the extreme-value limit process $Z_{\rm EV}(\ss)$ defined as in \eqref{eq:movingmaximum} (Section~\ref{subsec-est-2}), but with potentially different marginal distributions. We also assume that $F(\cdot; \tht_F)$ and $f(\cdot; \tht_F)$ are the marginal distribution and density functions, respectively, of the observed process $Y(\ss)$ and that $\tht_F$ is the vector of marginal parameters. In other words, the process $Y(\ss)$ has the same dependence properties (both in the bulk and the tails) as $Z(\ss)$, $\tilde{Z}(\ss)$ or $Z_{\rm EV}(\ss)$, but the marginal distributions are essentially arbitrary, thus allowing for a greater flexibility by modeling margins and dependence separately.}  

We may estimate marginal parameters non-parametrically in a first step using the empirical distribution (computed for each site separately, or for the entire pooled dataset). Alternatively, we can also estimate the vector of marginal parameters $\tht_F$ by maximizing the marginal (composite) log-likelihood function $\ell_F(\yy;\tht_F) = \sum_{i=1}^n \sum_{j=1}^d \log f(y_{ij};\tht_F)$, and we denote the respective estimate by $\widehat\tht_F$. Such an approach, which neglects spatial dependence to estimate marginal parameters, is known to be valid (i.e., yielding consistency and asymptotic normality of estimators) under mild regularity conditions \citep{Varin.etal:2011}. The data can then be transformed to the uniform ${\rm Unif}(0,1)$ scale using the probability integral transform. More precisely, pseudo-uniform scores $\{\uu_i = (u_{i1},\ldots,u_{id})^\top\}_{i=1}^n$ can be obtained by setting $u_{ij}=F(y_{ij};\widehat\tht_F)$.

In the next sections, we show how dependence parameters may be estimated from the pseudo-uniform scores $\{\uu_i\}_{i=1}^n$ by matching some suitable empirical and model-based summary statistics (for $Z(\ss)$ in (\ref{eq-convmodel}), and $\tilde Z(\ss)$ in (\ref{eq-convgauss})), or by exploiting composite likelihood inference (for $Z_{\mathrm{EV}}(\ss)$ in \eqref{eq:movingmaximum}).

%%%%%%%%%%%%%%%%%%%%%%%%%%%%%%%%%%%%%%%%%%%%%%%%
\subsection{Parameter estimation for the Cauchy convolution process model defined in \eqref{eq-convmodel}}
\label{subsec-est-1}
{Here, we assume that the pseudo-uniform scores $\{\uu_i\}_{i=1}^n$ obtained in Section~\ref{subsec-est-0} are driven by the copula stemming from the Cauchy convolution process $Z(\ss)$ in \eqref{eq-convmodel} with $W(\d\ss^*)\sim_{\text{i.i.d.}}\text{Cauchy}(\d\ss^*)$ using some parametric kernel function $k(\ss,\ss^*;\tht_K)$, where $\tht_K$ is the vector of dependence parameters controlling the kernel function. While the joint likelihood function for the process $Z(\ss)$ is not tractable in general, we can exploit its stochastic representation in \eqref{eq-convmodel} to estimate $\tht_K$.}

{We first back-transform the pseudo-uniform scores to the standard Cauchy scale. More precisely, we compute $\{\zz_i^* = (z_{i1}^*,\ldots,z_{id}^*)^\top\}_{i=1}^n$ with $z_{ij}^*=F^{-1}_{\CC}(u_{ij})$, where $F_{\CC}^{-1}(q) = \tan\{\pi(q - 0.5)\}$ is the quantile function of the standard Cauchy distribution. By stability of the Cauchy distribution, the marginal distribution of the process convolution $Z(\ss)$ in \eqref{eq-convmodel} are Cauchy with scale parameter 
$c(\ss;\tht_K) = \int_{\mathbb{R}^2} k(\ss,\ss^*;\tht_K)\d \ss^*$. 
This implies that $Z^*(\ss):=Z(\ss)/c(\ss;\tht_K)\sim{\rm Cauchy}(1)$ and that  $\{\zz_i^*\}_{i=1}^n$ can be treated as pseudo-observations from $Z^*(\ss)$. Moreover, we have
\begin{align*}
Z_{jk}^* &= Z^*(\ss_j)-Z^*(\ss_k)= {Z(\ss_j)\over c(\ss_j; \tht_K)} - {Z(\ss_k)\over c(\ss_k; \tht_K)} \\
&= \int_{\mathbb{R}^2}\{\zeta(\ss_j,\ss^*) - \zeta(\ss_k,\ss^*)\}  W(\d\ss^*) \ \sim \ \text{Cauchy}\{c_{jk}(\tht_K)\},
\end{align*}
where 
\begin{align}
\label{eq-cjk}
c_{jk}(\tht_K) &= \int_{\mathbb{R}^2}\left|\zeta(\ss_j,\ss^*) - \zeta(\ss_k,\ss^*)\right| \d\ss^*, \\ 
\zeta(\ss,\ss^*) &\equiv \zeta(\ss,\ss^*;\tht_K) = {k(\ss,\ss^*;\tht_K)\over \int_{\mathbb R^2}k(\ss,\ss^*;\tht_K)\d \ss^*}.\nonumber
\end{align}
For each pair of sites $\{\ss_j,\ss_k\}$, the scale parameter $c_{jk}$ may be estimated non-parametrically from the pseudo-observations $\{z_{ijk}^*=z_{ij}^*-z_{ik}^*\}_{i=1}^n$ by maximizing the corresponding Cauchy likelihood function. The maximum likelihood estimator $\widehat c_{jk}$ satisfies the equation $\sum_{i=1}^n{\widehat c_{jk}^2/(\widehat c_{jk}^2 + {z_{ijk}^*}^2)} = {n/2}$, whose positive root can be easily found using numerical routines. The estimator $\widehat c_{jk}$ is a consistent and asymptotically normal estimator of $c_{jk}$. Alternatively, the median of absolute values, ${\rm median}\{|z_{1jk}|,\ldots,|z_{njk}|\}$, may also be used as a more robust and faster-to-compute non-parametric estimator of $c_{jk}$, but which is about 20\% less efficient than the maximum likelihood estimator (in terms of the ratio of their variances); see \citet{Zhang2010}. To estimate the vector of parameters $\tht_K$, we can then use a least squares approach and compute
\begin{equation}
\label{eq:thtK-LS1}
\widehat\tht_K =\arg\min_{\tht_K} \sum_{j < k}\omega_{jk}\{c_{jk}(\tht_K) - \widehat c_{jk}\}^2,
\end{equation}
where $\omega_{kj}\geq0$ are some non-negative weights. While equal weights $\omega_{jk}=1$ are commonly chosen, binary weights specified according to the distance between sites, e.g., $\omega_{jk}=\ii(\|\ss_j-\ss_k\|\leq \delta_{\rm max})$ for some cut-off distance $\delta_{\rm max}>0$, may be helpful to reduce the computational burden and/or prioritize goodness-of-fit at small distances. The estimator in (\ref{eq:thtK-LS1}) is a special case of minimum distance estimators and therefore it is a consistent and asymptotically normal estimator of $\tht_K$ \citep{Millar1984}.}

%%%%%%%%%%%%%%%%%%%%%%%%%%%%%%%%%%%%%%%%%%%%%%%%
\subsection{Parameter estimation for the spatial mixture model extension defined in \eqref{eq-convgauss}}
\label{subsec-est-3}

{We now assume that the pseudo-uniform scores $\{\uu_i\}_{i=1}^n$ obtained in Section~\ref{subsec-est-0} are driven by the copula stemming from the extended model $\tilde{Z}(\ss)$ in \eqref{eq-convgauss}. In addition to the parametric kernel function $k(\ss,\ss^*;\tht_K)$ described by the vector of parameters $\tht_K$, we now need to estimate the correlation function $\rho_G(\delta;\tht_G)$ parametrized by a vector $\tht_G$, and the mixture parameter $\beta\geq0$.}

{Parameter estimation is now more tricky, but parameters can nevertheless be estimated in two steps by noticing that the copula of the limiting extreme-value process $Z_{\rm EV}(\ss)$ only depends on the kernel parameters $\tht_K$. In the first step, $\tht_K$ can thus be estimated by matching empirical and model-based estimates of the tail dependence coefficient for different pairs of sites. More precisely, let $\lambda_{jk}(\tht_K)$ be the tail dependence coefficient defined in \eqref{eq:lambdaU} corresponding to the pair of variables $\{\tilde{Z}(\ss_j),\tilde{Z}(\ss_k)\}$. On the one hand, $\lambda_{jk}$ can be estimated non-parametrically from the pseudo-uniform scores $\{(u_{ij},u_{ik})^\top\}_{i=1}^n$ for each pair of sites $\{\ss_j,\ss_k\}$, e.g., as
\begin{equation}\label{eq:lambdanonpar}
\widehat\lambda_{jk}={1\over n(1-u)}\sum_{i=1}^n\ii(u_{ij}>u,u_{ik}>u),
\end{equation}
where $u\approx 1$ is a high threshold on the uniform scale. The empirical estimator $\widehat\lambda_{jk}$ is consistent as $n\to\infty$ and $u\equiv u_n\to1$ such that $n(1-u_n)\to\infty$. Many other valid non-parametric estimators of the tail dependence coefficient exist. In particular, as the copula stemming from $\tilde{Z}(\ss)$ is reflection-symmetric, it would be possible to combine information from the lower and upper tails to estimate $\lambda_{jk}(\tht_K)$. Hereafter, we rely on an improved estimator proposed by \citet{Lee.Joe.ea2018}, which works well for small sample sizes. On the other hand, from Propositions~\ref{prop1-EVlim1} and \ref{prop2-EVlim1}, we have
\begin{align}
\lambda_{jk}(\tht_K) &= 2-\ell(1,1)\nonumber =2-\int_{\mathbb R^2} \max\{\zeta(\ss_j, \ss^*),\zeta(\ss_k, \ss^*)\}\d\ss^*\nonumber\\
&=1-\int_{\zeta(\ss_j,\ss^*) > \zeta(\ss_k,\ss^*)} \zeta(\ss_j, \ss^*)\d\ss^*+1-\int_{\zeta(\ss_k,\ss^*) > \zeta(\ss_j,\ss^*)} \zeta(\ss_k, \ss^*)\d\ss^*\nonumber\\
&=\int_{\zeta(\ss_j,\ss^*) \leq \zeta(\ss_k,\ss^*)} \zeta(\ss_j,\ss^*) \d\ss^* + \int_{\zeta(\ss_j,\ss^*) \geq \zeta(\ss_k,\ss^*)} \zeta(\ss_k,\ss^*) \d\ss^*,\label{eq:lambdamodel}
\end{align}
where $\ell(w_1,w_2)$ is the stable tail dependence function of $\{\tilde{Z}(\ss_j),\tilde{Z}(\ss_k)\}^\top$, and $\zeta(\ss,\ss^*) \equiv \zeta(\ss,\ss^*;\tht_K) = {k(\ss,\ss^*;\tht_K)/\int_{\mathbb R^2}k(\ss,\ss^*;\tht_K)\d \ss^*}$. The coefficient $\lambda_{jk}(\tht_K)$ can be expressed in simple form as a function of $\tht_K$ for many parametric families of kernels. We consider one example in the simulation study reported in Section~\ref{sec-simstudy}. The parameter $\tht_K$ can then be estimated by least squares as follows:
\begin{equation}\label{eq:thtK-LS2}
\widehat\tht_K = \arg\min_{\tht_K}\sum_{j<k}\omega_{jk}\{\lambda_{jk}(\tht_K) - \widehat\lambda_{jk}\}^2,
\end{equation}
where $\widehat\lambda_{jk}$ are the empirical estimates from \eqref{eq:lambdanonpar} and $\omega_{jk}\geq0$ are some non-negative weights as in \eqref{eq:thtK-LS1}. Notice that this approach based on the tail dependence coefficient could also be applied to the Cauchy process $Z(\ss)$, as it shares the same extremal dependence structure, but the least squares estimator \eqref{eq:thtK-LS1} is more efficient than \eqref{eq:thtK-LS2} as it uses information from both the bulk and the tails.}

{After having estimated the kernel parameters $\tht_K$, we now need to estimate the parameters $\tht_G$ of the correlation function $\rho_G(\delta) = \rho_G(\delta; \tht_G)$ and the parameter $\beta\geq0$ in \eqref{eq-convgauss}. Let $F(\cdot; \gamma, \beta)$ be the CDF of the variable $\gamma W + \beta Z$ where $W$ and $Z$ are independent random variables following the standard Cauchy and Gaussian distributions, respectively. The distribution function of this sum of variables may be expressed in integral form as 
\begin{equation}
\label{eq:CauchyConvNormalCDF}
F(w; \gamma, \beta)={1\over 2}+{1\over \pi}\int_{\mathbb R}\arctan\{(w-\beta z)/\gamma\}\phi(z)\d z,
\end{equation}
where $\phi(\cdot)$ is the standard normal density function. Numerical integration can be used to quickly and accurately evaluate \eqref{eq:CauchyConvNormalCDF}. The corresponding density function can be computed by differentiating \eqref{eq:CauchyConvNormalCDF} under the integral sign. Fixing the value of $\beta = \beta_0\geq0$, we then back-transform the pseudo-uniform scores to the scale of $F(\cdot; 1, \beta_0)$ as $\{\tilde{\zz}_i=(\tilde{z}_{i1},\ldots,\tilde{z}_{id})^\top\}_{i=1}^n$, where $\tilde{z}_{ij}=F^{-1}(u_{ij};1,\beta_0)$. If $\beta_0$ is the true value of $\beta$, then the vectors $\{\tilde{\zz}_i\}_{i=1}^n$ can be considered as pseudo-observations from the process $\tilde{Z}(\ss)$ in \eqref{eq-convgauss}. Now, notice that for any two sites $\{\ss_j,\ss_k\}$, we have
\begin{align*}
\tilde{Z}^+_{jk} &= \tilde{Z}(\ss_j) + \tilde{Z}(\ss_k)  \sim F\{\cdot; 2, \beta^+_{jk}(\tht_G)\}, \\
\tilde{Z}^-_{jk} &= \tilde{Z}(\ss_j) - \tilde{Z}(\ss_k)  \sim F\{\cdot; c_{jk}(\tht_K), \beta^-_{jk}(\tht_G)\},
\end{align*}
where $c_{jk}(\tht_K)$ is given in (\ref{eq-cjk}) and $\beta^+_{jk}(\tht_G) = \beta_0 \{2+2\rho_G(\delta_{jk}; \tht_G)\}^{1/2}$, $\beta^-_{jk}(\tht_G) = \beta_0 \{2-2\rho_G(\delta_{jk};\tht_G)\}^{1/2}$, $\delta_{jk} = \|\ss_j - \ss_k\|$. To obtain empirical estimates of $\tht_G$, we first obtain non-parametric estimates $\widehat\beta_{jk}^+$ of $\beta_{jk}^+(\tht_G)$ by assuming that $\tilde{Z}^+_{jk}\sim F\{\cdot;2,\beta_{jk}^+(\tht_G)\}$ and maximizing the corresponding likelihood function using the pseudo-observations $\{\tilde{z}^+_{jk}=\tilde{z}_{ij}+\tilde{z}_{ik}\}_{i=1}^n$. We use the same maximum likelihood approach to get non-parametric estimates $\widehat\beta_{jk}^-$ of $\beta_{jk}^-(\tht_G)$, but now assuming that $\tilde{Z}^-_{jk}\sim F\{\cdot;c_{jk}(\widehat\tht_K),\beta_{jk}^-(\tht_G)\}$ with $\widehat\tht_K$ obtained in \eqref{eq:thtK-LS2} and using the pseudo-observations $\{\tilde{z}^-_{jk}=\tilde{z}_{ij}-\tilde{z}_{ik}\}_{i=1}^n$ instead. The vector of parameters $\tht_G$ and the parameter $\beta\geq0$ can then be jointly estimated by least squares as
\begin{equation}\label{eq:thtG-beta-LS}
(\widehat\tht_G^\top,\widehat\beta)^\top = \arg\min_{(\tht_G^\top,\beta_0)^\top}\sum_{j < k}\omega_{jk}^+\{\beta_{jk}^+(\tht_G) - \widehat\beta^+_{jk}\}^2 + \sum_{j < k}\omega_{jk}^-\{\beta_{jk}^-(\tht_G) - \widehat\beta^-_{jk}\}^2,
\end{equation}
where $\omega_{jk}^+\geq0$ and $\omega_{jk}^-\geq0$ are some non-negative weights associated to each pair of sites $\{\ss_j,\ss_k\}$ as in \eqref{eq:thtK-LS1} and \eqref{eq:thtK-LS2}. In practice, the minimization in \eqref{eq:thtG-beta-LS} can be performed over $\tht_G$ for a grid of fixed values $\beta=\beta_0\geq0$, and then the value $\beta$ can be selected in a second step as the one that provides the lowest objective function overall. The estimators in (\ref{eq:thtK-LS2}) and (\ref{eq:thtG-beta-LS}) are again special cases of minimum distance estimators and they are therefore consistent and asymptotically normal if $\widehat\lambda_{j,k}$ and $\widehat\beta^+_{jk}, \widehat\beta^-_{jk}$ are consistent and asymptotically normal estimators of $\lambda_{j,k}$ and $\beta^+_{jk}(\tht_G), \beta^-_{jk}(\tht_G)$, respectively \citep{Millar1984}.

\subsection{Parameter estimation for the extreme-value model $Z_{\rm EV}(s)$ in~\eqref{eq:movingmaximum}}
\label{subsec-est-2}

{We now assume that the pseudo-uniform scores $\{\uu_i\}_{i=1}^n$ obtained in Section~\ref{subsec-est-0} are driven by the copula stemming from the extreme-value process $Z_{\rm EV}(\ss)$ in \eqref{eq:movingmaximum} with parametric kernel function $k(\ss,\ss^*;\tht_K)$, obtained as limit of the Cauchy convolution process \eqref{eq-convmodel} or the spatial mixture model extension \eqref{eq-convgauss}. Such an extreme-value copula has the form \eqref{eq:stabletail}. When the stable tail dependence function $\ell$ given in Proposition~\ref{prop1-EVlim1} has a simple explicit form, a pairwise (composite) likelihood approach may be used to estimate the dependence parameters $\tht_K$ \citep{Lindsay.1998,Padoan.etal:2010,Varin.etal:2011}. Let $C_{{\rm EV},jk}$ be the bivariate extreme-value copula restricted to the pair of sites $\{\ss_j,\ss_k\}$, and $c_{{\rm EV},jk}$ be the corresponding density function. More precisely, we have $C_{{\rm EV},jk}(u_j,u_k)=\exp\{-\ell_{jk}(-\log u_j, -\log u_k; \tht_K)\}$, where $\ell_{jk}(w_j,w_k)=\ell(w_j\ee_j+w_k\ee_k)$ (with $\ee_j$ the $j$-th canonical basis vector of $\mathbb R^d$) is the $(j,k)$-th margin of $\ell(w_1,\ldots,w_d)$, and 
	\begin{align*}
	c_{{\rm EV},jk}(u_j,u_k)&=(u_ju_k)^{-1}\exp\{-\ell_{jk}(-\log u_j, -\log u_k; \tht_K)\}\\&\times\{\partial_1 \ell_{jk}(-\log u_j, -\log u_k; \tht_K) \partial_2 \ell_{jk}(-\log u_j, -\log u_k; \tht_K)\\&\hspace{0.5cm}-\partial_{12} \ell_{jk}(-\log u_j, -\log u_k; \tht_K)\}, 
	\end{align*}
	where $\partial_1$ means differentiation with respect to the first argument, and so forth. The parameter vector $\tht_K$ may then be estimated by maximizing a pairwise log-likelihood function as follows:
	\begin{equation}\label{eq:MPLE}
	\widehat\tht_K=\arg\max_{\tht_K}\sum_{j < k} \omega_{jk}\log \{c_{{\rm EV},jk}(u_{ij}, u_{ik}; \tht_K)\},
	\end{equation}
	where $\omega_{jk}\geq0$ are some non-negative weights. %as in \eqref{eq:thtK-LS1}, \eqref{eq:thtK-LS2} and \eqref{eq:thtG-beta-LS}. 
	Under mild regularity conditions, the maximum pairwise likelihood estimator \eqref{eq:MPLE} is consistent and asymptotically normal, but with some loss of efficiency compared to the usual maximum likelihood estimator since it only uses information contained in pairs of variables; see, e.g., \citet{Padoan.etal:2010} and \citet{Huser:2013}, Chapter~3. Notice that when the stable tail dependence function $\ell$ is not tractable or is difficult to compute, it is also possible to estimate $\tht_K$ by matching empirical and model-based coefficient of tail dependence given by $\lambda_{jk}=2-\ell_{jk}(1,1)$. This approach is similar to the least squares estimator \eqref{eq:thtK-LS2} in the main manuscript, except that the coefficients $\lambda_{jk}$ may be estimated from the full dataset of maxima instead of just from the tail. Various estimators have been proposed to estimate $\lambda_{jk}$ non-parametrically, and in our simulation study we relied on an estimator of the so-called Pickands dependence function; see for example \cite{Genest.Segers2009}.}
	
%%%%%%%%%%%%%%%%%%%%%%%%%%%%%%%%%%%%%%%%%%%%%%%%
%%%%%%%%%%%%%%%%%%%%%%%%%%%%%%%%%%%%%%%%%%%%%%%%
%%%%%%%%%%%%%%%%%%%%%%%%%%%%%%%%%%%%%%%%%%%%%%%%

\section{Numerical experiments}
\label{sec-numericalexp}

%%%%%%%%%%%%%%%%%%%%%%%%%%%%%%%%%%%%%%%%%%%%%%%%
\subsection{Simulation study}
\label{sec-simstudy}

{We now perform a simulation study to verify the performance of the inference schemes proposed in Sections~\ref{subsec-est-1}, \ref{subsec-est-3}, and \ref{subsec-est-2}, before illustrating our proposed methodology by application to real temperature data in Section~\ref{sec-empstudy}. For estimating kernel parameters in the mixture process $\tilde{Z}(\ss)$, we rely on the least squares approach \eqref{eq:thtK-LS2} and we use on a nonparametric estimator of the tail dependence coefficient proposed by \citet{Lee.Joe.ea2018} that was found to be efficient and accurate in small samples. For the extreme-value process $Z_{\rm EV}(\ss)$, we similarly estimate kernel parameters by matching empirical and model-based coefficients of tail dependence given now by $\lambda_{jk}=2-\ell_{jk}(1,1)$. This approach is computationally tractable in high dimensions even if the stable tail dependence function $\ell$ has no simple form, and it is akin to the least squares estimator \eqref{eq:thtK-LS2} except that the coefficients $\lambda_{jk}$ can be estimated from the full dataset instead of just from the tail. Various nonparametric estimators have been proposed for $\lambda_{jk}$, and we here rely on an estimator of the so-called Pickands dependence function; see \citet{Genest.Segers2009}. } 

{We simulate datasets comprised of $n=100,200,500$ replicates of the Cauchy convolution process $Z(\ss)$ in \eqref{eq-convmodel} for some kernel function $k(\ss,\ss^*;\tht_K)$, its spatial mixture process extension $\tilde{Z}(\ss)$ in \eqref{eq-convgauss}, and the extreme-value limit $Z_{\rm EV}(\ss)$ in \eqref{eq:movingmaximum}, at $d=9,25,100$ sites on the regular grid $\left\{(j/(m+1), k/(m+1))^\top\right\}_{j,k=1}^m\subset \mathbb [0,1]^2$, for $m=3, 5, 10$. To illustrate the methods, we consider here the compactly supported kernel function $k(\ss,\ss^*;\tht_K) = \left(1 - \|\ss-\ss^*\|/r\right)^{\eta}_+$ with true kernel parameters chosen as $\tht_K = (\eta, r)^\top=(1, 0.25)^\top$. For the spatial mixture process $\tilde{Z}(\ss)$ in \eqref{eq-convgauss}, we additionally specify the correlation function of the Gaussian process to be $\rho_G(\delta;\theta_G) = \exp(-\theta_{G}\delta)$ with rate $\theta_{G} = 1$, and the mixture parameter is fixed to $\beta = 2$.}

{Following the notation from Section \ref{subsec-est-1}, we need to find the theoretical expression of the scale parameter $c_{jk}(\tht_K)$ in \eqref{eq-cjk}. By symmetry, straightforward calculations yield
\begin{align*}
c_{jk}(\tht_K) &= \int_{\mathbb{R}^2}\left|\zeta(\ss_j,\ss^*) - \zeta(\ss_k,\ss^*)\right| \d\ss^*\\
&= 2\int_{\zeta(\ss_j,\ss^*) > \zeta(\ss_k,\ss^*)} \zeta(\ss_j,\ss^*) \d\ss^* + 2\int_{\zeta(\ss_j,\ss^*) < \zeta(\ss_k,\ss^*)} \zeta(\ss_k,\ss^*) \d\ss^* - 2\\
&= {4\over c^*(\tht_K)}\int_{\|\ss^*-\ss_k\| \leq \min\{r, \|\ss_j-\ss^*\|\}} \left(1 - {\|\ss_k-\ss^*\|\over r}\right)^{\eta} \d\ss^* - 2,
\end{align*}
where $\zeta(\ss,\ss^*) = {k(\ss,\ss^*;\tht_K)/\int_{\mathbb R^2}k(\ss,\ss^*;\tht_K)\d \ss^*}$ and the normalizing factor is equal to $c^*(\tht_K) = \int_{\mathbb R^2} k(\ss, \ss^*; \tht_K) \d\ss^* = 2\pi r^2/\{(\eta+1)(\eta+2)\}$. 
After a change of variables, we find that 
\begin{align*}
c_{jk}(\tht_K) &= {4r^2\over c^*(\tht_K)}\int_{\pi}^{2\pi} \left\{{g_{jk}(\phi; r)^{\eta+1}\over \eta+1} - {g_{jk}(\phi; r)^{\eta+2}\over\eta+2}\right\}\d\phi\\
&= {2(\eta+1)(\eta+2)\over \pi}\int_{\pi}^{2\pi} \left\{{g_{jk}(\phi; r)^{\eta+1}\over \eta+1} - {g_{jk}(\phi; r)^{\eta+2}\over\eta+2}\right\}\d\phi,
\end{align*}
where $g_{jk}(\phi;r) = 1 - \max\left\{0, 1 + \|\ss_j-\ss_k\|/(2r\sin\phi)\right\}$. This integral can be easily computed numerically. Furthermore, from \eqref{eq:lambdamodel}, we can also deduce that $\lambda_{jk}(\tht_K) = 1 - c_{jk}(\tht_K)/2$. 

For each simulated dataset we use the Student-$t$ marginal distribution with four degrees of freedom for the processes $Z(\ss)$ and $\tilde{Z}(\ss)$, and the Fr\'echet marginal distribution with the shape parameter 4 for the process $Z_{\rm EV}(\ss)$. Degrees of freedom and shape parameters are estimated in a first step using the marginal likelihood approach described in Section \ref{subsec-est-0}. Dependence (i.e., copula) parameters are then estimated in a second step based on a pairwise least squares inference approach (recall Sections~\ref{subsec-est-1}, \ref{subsec-est-3}, and \ref{subsec-est-2}). Here, we set the weights to $\omega_{jk}^+=\omega_{jk}^-=1$ for all pairs of sites $\{\ss_j,\ss_k\}$ with $||\ss_j - \ss_k|| < \delta_{\rm max}$ in \eqref{eq:thtK-LS1}, \eqref{eq:thtK-LS2}, and \eqref{eq:thtG-beta-LS}, where we use $\delta_{\rm max} = 0.4$ for $d = 9, 25$ and $\delta_{\rm max} = 0.25$ for $d=100$. Other weights are set to zero, so we only include pairs of locations at small distances to improve the accuracy of estimates and make computations faster, especially for $d=100$.  More specifically, the values of $\delta_{\rm max}$ are chosen to keep 20, 150, and 790 close-by pairs for inference, i.e., about 56\%, 50\%, and 16\% of all pairs for $d=9,25,100$, respectively, in order to achieve a reasonable trade-off between computational and statistical efficiency.
%though this may not be the optimal choice. 
Parameters to be estimated are the kernel parameters $\tht_K=(\eta,r)^\top$ for $Z(\ss)$ and $Z_{\rm EV}(\ss)$, and $(\eta,r,\theta_G,\beta)^\top$ for the spatial mixture process $\tilde{Z}(\ss)$. To assess the estimators' performance, we repeat the simulations $N=500$ times and compute the root mean squared errors (RMSE) for all estimated parameters. We also compute 
%\begin{eqnarray*}
$\Delta_{\max} = {1\over N}\sum_{i=1}^N \max_{x \in (0,1)}|k(\ss_0,\ss(x); \tht_K) - k(\ss_0,\ss(x); \widehat\tht_{K,i})|$, and 
$\Delta_{\mathrm{avg}} = {1\over N}\sum_{i=1}^N \int_0^1 |k(\ss_0,\ss(x); \tht_K) - k(\ss_0,\ss(x); \widehat\tht_{K,i})|\d x$, 
%\end{eqnarray*}
where $\ss_0 = (0,0)^\top$ and $\ss(x) = (0,x)^\top$, while $\tht_K=(1,0.25)^\top$ denotes the true kernel parameters and $\widehat\tht_{K,i}$ is its estimate for the for $i$-th simulation ($i=1,\ldots,N$). Hence, $\Delta_{\max}$ and $\Delta_{\mathrm{avg}}$ represent the maximum and mean integrated absolute differences between the true kernel and its estimate along a horizontal segment passing through the origin, averaged across the $N=500$ simulations.}

Table \ref{tab1-RMSER} reports the results. As expected, estimates are more accurate with larger sample sizes, as shown by significantly smaller RMSE and $(\Delta_{\max},\Delta_{\mathrm{avg}})$ values as $n$ increases. Moreover, using data from more locations (i.e., increasing $d$) can further improve the accuracy of parameter estimates.

\begin{table}[t!]
	\caption{\footnotesize {Performance metrics (i.e., RMSE of estimated parameters and $(\Delta_{\max},\Delta_{\mathrm{avg}}))$ calculated for the least squares inference schemes detailed in Sections~\ref{subsec-est-1}, \ref{subsec-est-3}, and the appendix for data simulated from the processes $Z(\ss)$ (top panel), $\tilde{Z}(\ss)$ (middle panel), and $Z_{\rm EV}(\ss)$ (bottom panel), respectively. The true kernel function is chosen as $k(\ss,\ss^*;\tht_K) = \left(1 - \|\ss-\ss^*\|/r\right)^{\eta}_+$, with parameters $\tht_K = (\eta, r)^\top=(1, 0.25)^\top$. For the spatial mixture process $\tilde{Z}(\ss)$, the correlation function of the underlying Gaussian process is $\rho_G(\delta;\theta_G) = \exp(-\theta_{G}\delta)$ with $\theta_{G} = 1$, and the mixture parameter is $\beta = 2$. Datasets are simulated on a regular grid at $d=9,25,100$ locations, with $n=100,200,500$ independent replicates.}}
	\label{tab1-RMSER}
	\def~{\hphantom{0}}
	\begin{center}
	{Data simulated from the Cauchy process $Z(\ss)$ in \eqref{eq-convmodel}, with inference based on Section~\ref{subsec-est-1}.}
	
	\vspace{4pt}
	
	{\begin{tabular}{c|>{\centering\arraybackslash}p{4cm}>{\centering\arraybackslash}p{4cm}>{\centering\arraybackslash}p{2.5cm}}
		&\multicolumn{3}{c}{RMSE for $\tht_K^\top = (\eta, r)$ (top) and $(\Delta_{\max}, \Delta_{\mathrm{avg}})$ (bottom)}\\
	        Sample size & $d=9$ & $d=25$ & $d=100$ \\ 	
	        \hline
	        $n=100$&(2.49, 0.16)&(1.62,0.11)&(0.30,0.02)\\
	        &(0.21,0.03)&(0.14,0.02)&(0.07,0.01)\\ [3pt]
	        $n=200$&(1.86,0.12)&(0.69,0.05)&(0.22,0.02)\\
	        &(0.16,0.03)&(0.08,0.01)&(0.05,0.01)\\ [3pt]
	        $n=500$&(0.85,0.05)&(0.26,0.02)&(0.13,0.01)\\
	        &(0.11,0.02)&(0.04,0.01)&(0.03,0.00)
	\end{tabular}}
	
	\vspace{14pt}
	
	{Data simulated from the spatial mixture process $\tilde{Z}(\ss)$ in \eqref{eq-convgauss}, with inference based on Section~\ref{subsec-est-3}.}
	
	\vspace{4pt}
	
	{\begin{tabular}{c|>{\centering\arraybackslash}p{4cm}>{\centering\arraybackslash}p{4cm}>{\centering\arraybackslash}p{2.5cm}}
		&\multicolumn{3}{c}{RMSE for $(\eta, r,\theta_G,\beta)$ (top) and $(\Delta_{\max}, \Delta_{\mathrm{avg}})$ (bottom)}\\
	        Sample size & $d=9$ & $d=25$ & $d=100$ \\ 	
	        \hline
	        $n=100$&(1.77,0.14,0.76,0.86)&(1.02,0.08,0.67,0.75)&(1.00,0.08,0.52,0.74)\\
	        &(0.36,0.05)&(0.15,0.02)&(0.11,0.02)\\[3pt]
	        $n=200$&(1.80,0.14,0.73,0.79)&(0.78,0.06,0.65,0.73)&(0.63,0.04,0.46,0.64)\\
	        &(0.33,0.05)&(0.13,0.02)&(0.09,0.01)\\[3pt] 
	        $n=500$&(1.52,0.11,0.56,0.61)&(0.45,0.03,0.51,0.55)&(0.52,0.03,0.30,0.40)\\
	        &(0.21,0.03)&(0.09,0.01)&(0.08,0.01)
	\end{tabular}}
	
	\vspace{14pt}
	
	{Data simulated from the extreme-value process $Z_{\rm EV}(\ss)$ in \eqref{eq:movingmaximum}, with inference based on a least squares approach based on pairwise tail dependence coefficients (see Appendix).}%Section~\ref{subsec-est-2}.}
	
	\vspace{4pt}
	
	{\begin{tabular}{c|>{\centering\arraybackslash}p{4cm}>{\centering\arraybackslash}p{4cm}>{\centering\arraybackslash}p{2.5cm}}
		&\multicolumn{3}{c}{RMSE for $\tht_K^\top = (\eta, r)$ (top) and $(\Delta_{\max}, \Delta_{\mathrm{avg}})$ (bottom)}\\
	        Sample size & $d=9$ & $d=25$ & $d=100$ \\ 	
	        \hline
	        $n=100$&(1.63,0.12)&(0.65,0.05)&(0.31,0.03)\\
	        &(0.26,0.04)&(0.10,0.02)&(0.06,0.01)\\ [3pt]
	        $n=200$&(1.30,0.09)&(0.43,0.03)&(0.22,0.02)\\
	        &(0.17,0.03)&(0.08,0.01)&(0.04,0.01)\\ [3pt]
	        $n=500$&(0.82,0.06)&(0.28,0.02)&(0.13,0.01)\\
	        &(0.12,0.02)&(0.05,0.01)&(0.03,0.00)
	\end{tabular}}
	\end{center}
\end{table}

For small $n$ and $d$, RMSE values are quite large due to kernel parameters being more tricky to identify. Similar kernels (hence, dependence structures) may be obtained for different combinations of parameters $\tht_K = (\eta, r)^\top$, and thus the effects of $\eta$ (shape) and $\rho$ (dependence range) are difficult to distinguish, especially in low sample sizes. Figure \ref{fig3kernels} shows the estimated kernel profiles along the $x$-axis, i.e., $k(\ss_0,\ss(x), \widehat\tht_{K,i})$ with $0 < x < 1$, for each simulated dataset of $Z(\ss)$ and $Z_{\rm EV}(\ss)$ with $d = 100$. The true kernel appears to be nevertheless very well estimated, even when the sample size is not very large. Similar results are obtained with other sets of parameters and with different kernel functions.

\begin{figure}[t!]
	\begin{center}
		\vspace{-1cm}
		\includegraphics[width=0.9\linewidth]{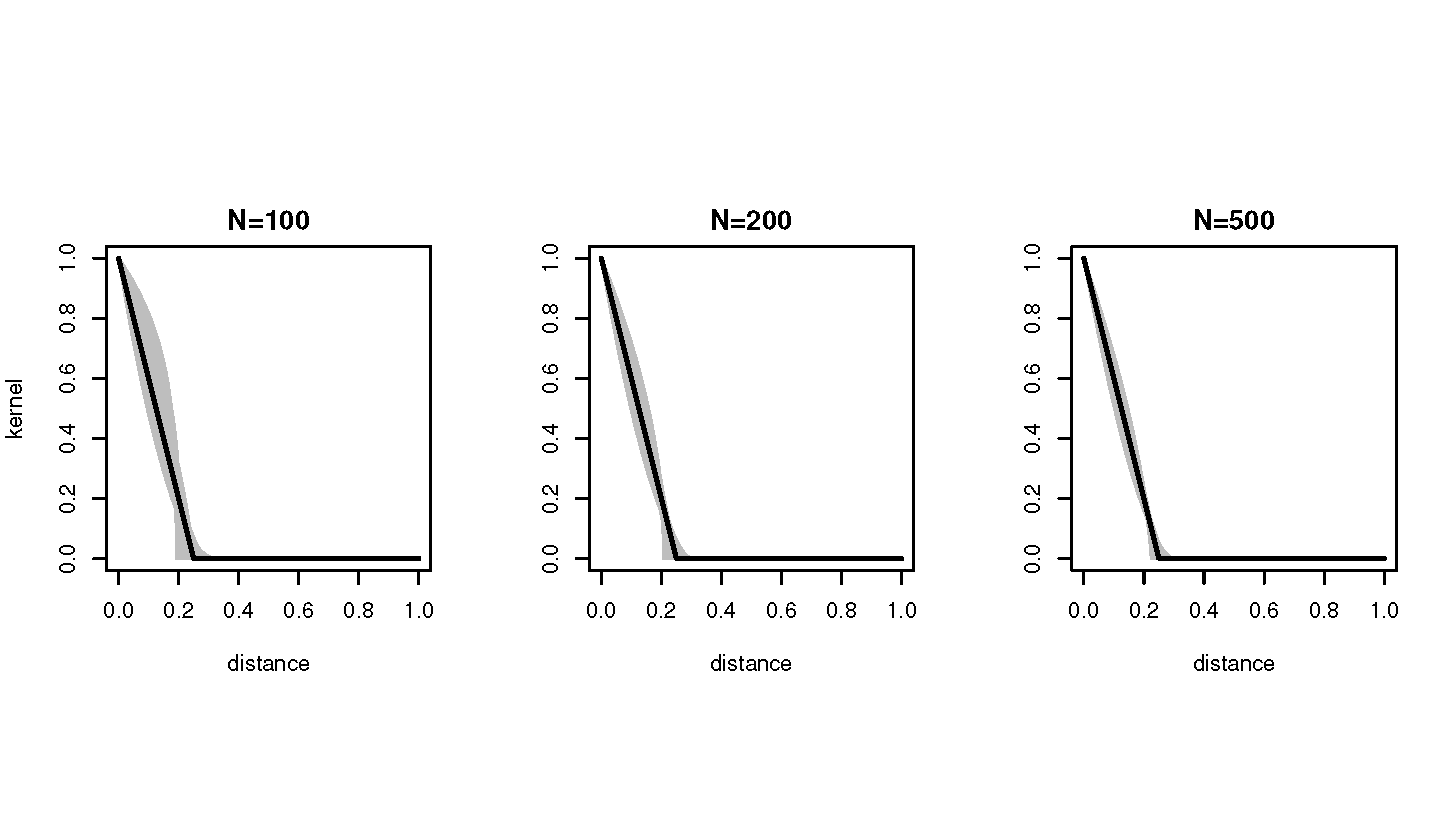}\vspace{-2cm}
		\includegraphics[width=0.9\linewidth]{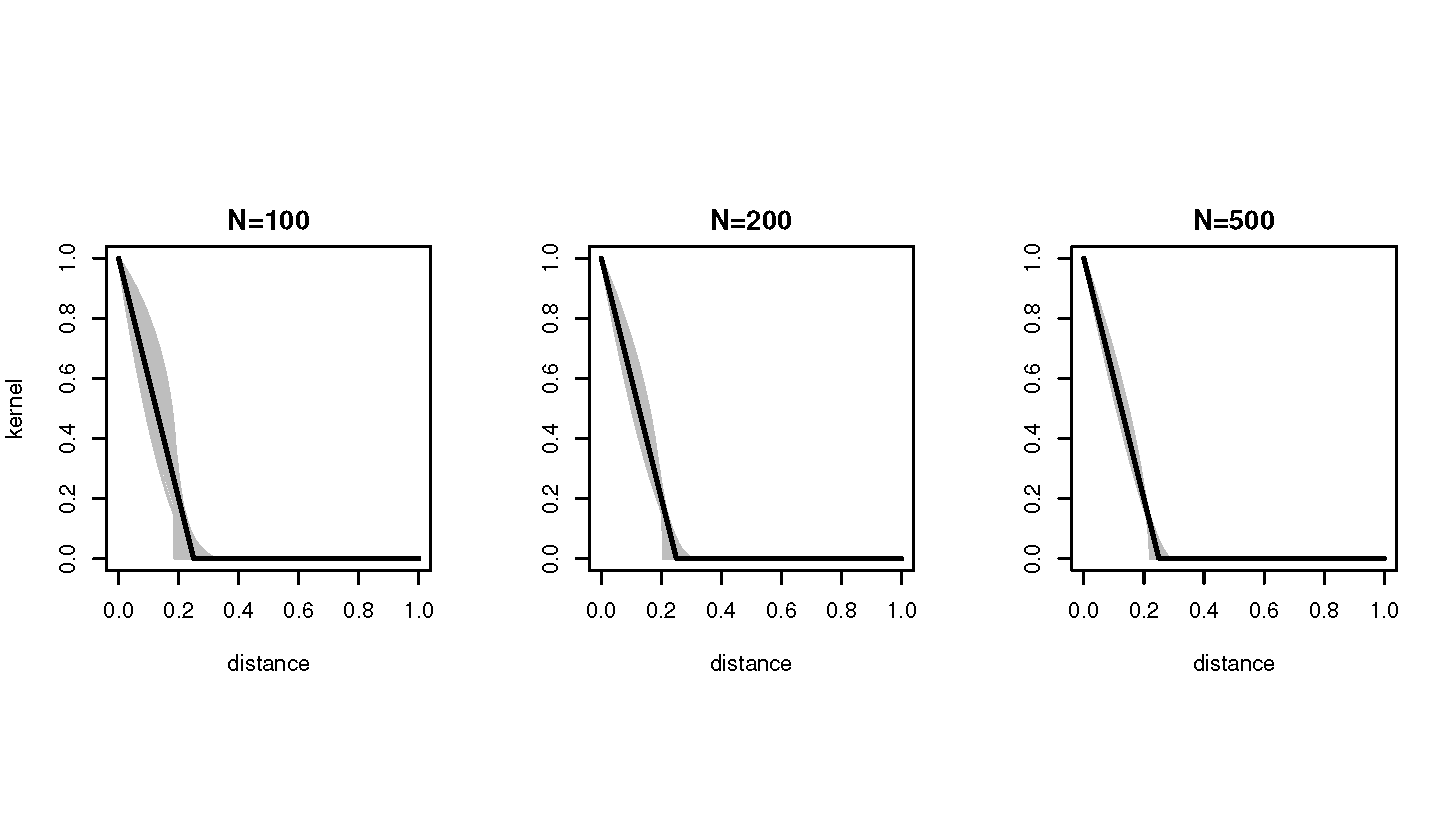}\vspace{-1cm}
		\caption{{\footnotesize {Estimated kernel profiles $k(\ss_0, \ss(x); \widehat\tht_{K,i})$, $0 < x < 1$, $i=1,\ldots,N$ (light gray lines), based on $N=500$ simulated processes $Z(\ss)$ (top row) and extreme-value limit $Z_{\rm EV}(\ss)$ (bottom row), with $n=100$ (left column), $200$ (middle column) and $500$ (right column) independent replicates. Black thick solid lines show the true kernels $k(\ss_0, \ss(x; \tht_K))$ with $\tht_K = (1, 0.25)^\top$.}}}
		\label{fig3kernels}
	\end{center}
\end{figure}

%%%%%%%%%%%%%%%%%%%%%%%%%%%%%%%%%%%%%%%%%%%%%%%%
\subsection{Temperature data application}
\label{sec-empstudy}

We now illustrate the proposed methodology to analyze temperature data from the state of Oklahoma, United States. Here, we use daily temperature averages measured at 97 monitoring stations at maximum distance $531$km. %20 stations, namely \texttt{Acme}, \texttt{Ada}, \texttt{Ardmore2}, \texttt{Byars}, \texttt{Centrahoma}, \texttt{Chikasha}, \texttt{Durant}, \texttt{Fittstown}, \texttt{Ketchum Ranch}, \texttt{Lane}, \texttt{Madill}, \texttt{Minco}, \texttt{Newport}, \texttt{Norman}, \texttt{Pauls Valley}, \texttt{Ringling}, \texttt{Shawnee}, \texttt{Tishomingo}, \texttt{Washington}, and \texttt{Waurika}. 
The time period considered here is the year 2018, which contains $n=366$ days in total. The dataset can be freely downloaded from the website \texttt{mesonet.org}. After removing the obvious seasonal component, we then fit an \texttt{AR(2)} model to account for temporal dependence, and we fit the skew-$t$ distribution jointly to the residuals using the marginal likelihood approach described in Section \ref{subsec-est-0}. Finally, we transform the residuals to the ${\rm Unif}(0,1)$ scale using the estimated marginal distribution functions. To explore the dependence features of the data, the left panel of Figure~\ref{fig0empstudy} displays bivariate scatter plots of normal scores (i.e., residuals further transformed to the standard normal distribution) for some selected pairs of stations.}  
\begin{figure}[t!]
	\begin{center}
		\includegraphics[width=0.4\linewidth]{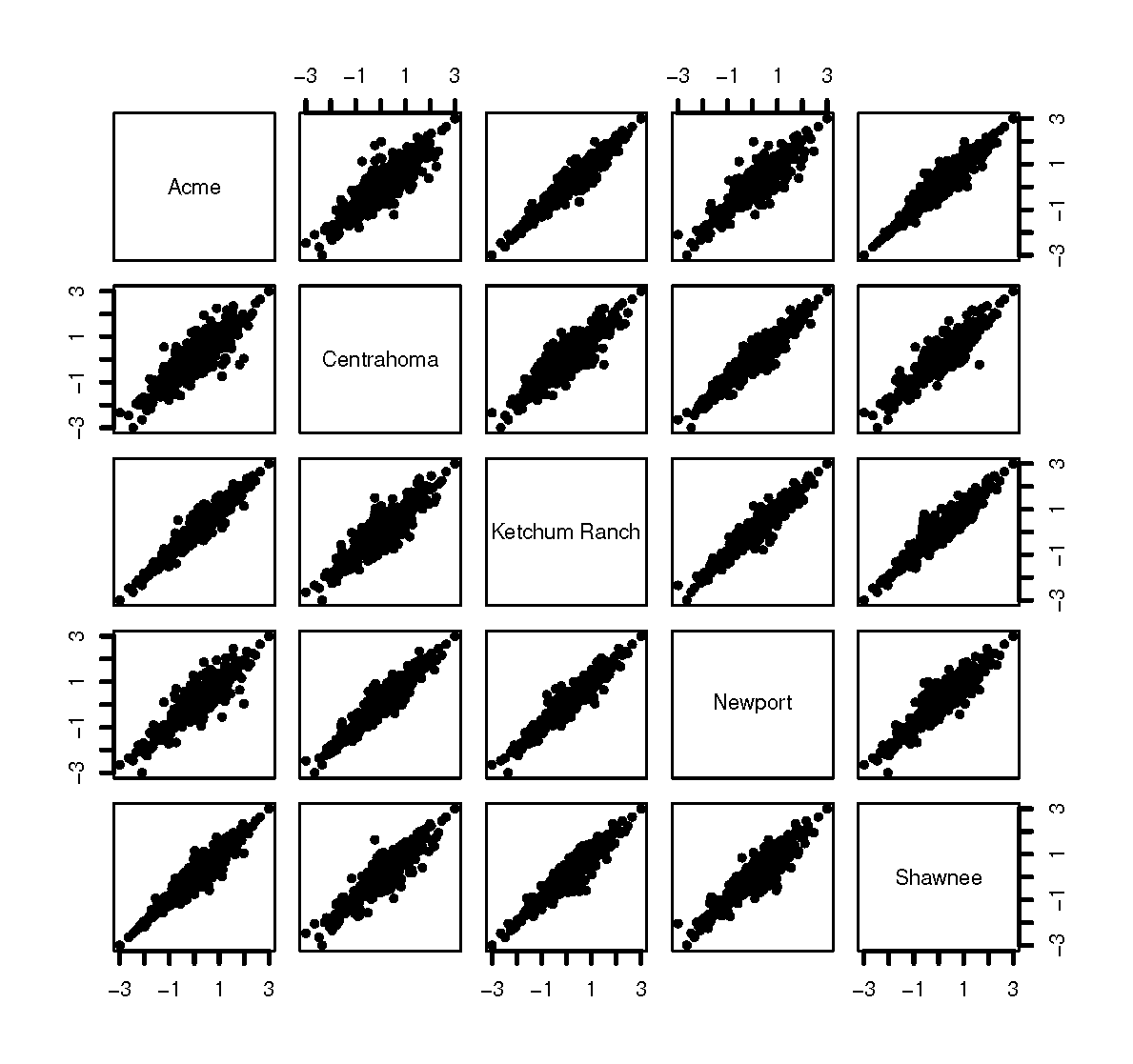}\hspace{-0.5cm}
		\includegraphics[width=0.4\linewidth]{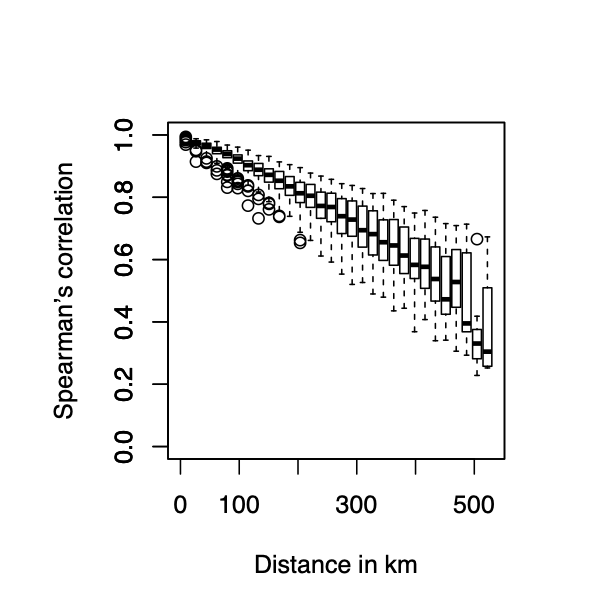}
		\caption{{\footnotesize {\emph{Left:} Bivariate scatter plots of normal scores based on the temperature data residuals (obtained from preliminary marginal fits) for some selected pairs of stations. \emph{Right:} Spearmans's correlation estimates for all pairs of stations, plotted as a function of distance between stations.}}}
		\label{fig0empstudy}
	\end{center}
\end{figure}
{The sharp and non-elliptical tails that can be seen in all bivariate scatter plots indicate clear evidence of non-Gaussianity and strong tail dependence. Furthermore, these diagnostics do not reveal any significant tail asymmetry. Therefore, while a (transformed) Gaussian process would clearly not be suitable to model the data due to its very weak form of dependence in the tails, our proposed Cauchy convolution process provides a more adequate alternative. Furthermore, the right panel of Figure~\ref{fig0empstudy} shows Spearman's correlation estimates for all pairs of stations, plotted as a function of distance between stations. Although these empirical estimates appear quite dispersed, this plot reveals that the behavior of Spearman's correlation function is approximately linear near the origin, which suggests that the observed process is quite smooth and that our Cauchy convolution model should fit well.}

\begin{figure}[t!]
	\begin{center}
		\includegraphics[width=0.33\linewidth]{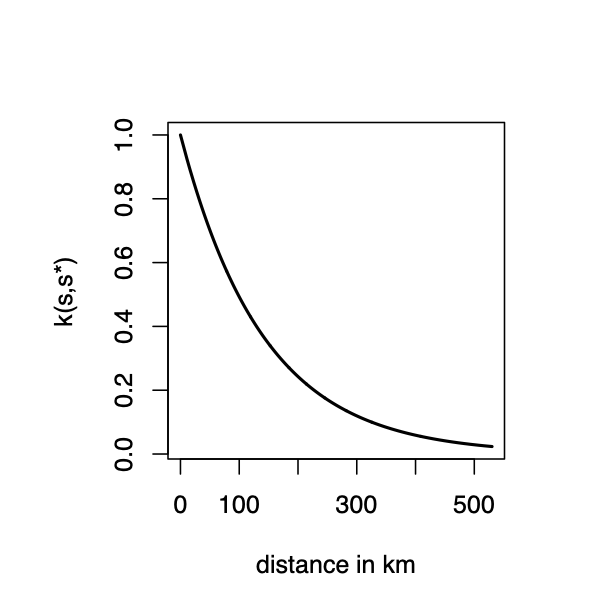}\hspace{-12mm}
		\includegraphics[width=0.33\linewidth]{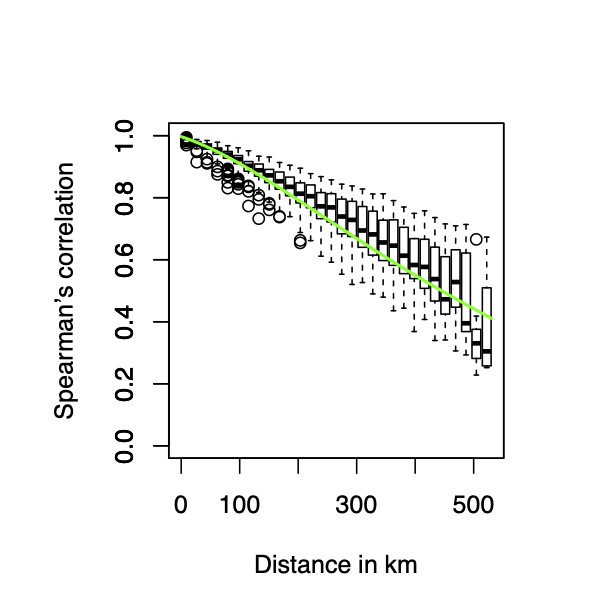}\hspace{-12mm}
		\includegraphics[width=0.33\linewidth]{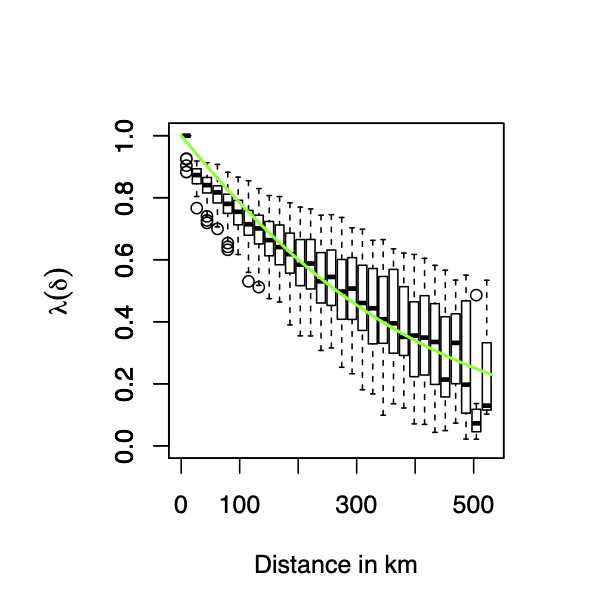}
		\caption{{\footnotesize {\emph{Left:} Fitted kernel function $k(\ss,\ss^*;\widehat\tht_K) = (1-\|\ss-\ss^*\|/\widehat r)_+^{\widehat\eta}$ with $\widehat\tht_K =(\widehat\eta,\widehat r)^\top = (8985, 1270)^\top$. \emph{Middle:} Empirical estimates of Spearman's correlation for all pairs of stations (boxplots) and fitted Spearman's correlation function $S_\rho(\delta)$ (solid green light), plotted as a function of distance $\delta$ between stations. \emph{Right:} Empirical upper tail dependence coefficient estimates for all pairs of stations (boxplots) and fitted tail dependence function $\lambda(\delta)$ (solid green light), plotted as a function of distance $\delta$ between stations.}}}
		\label{fig2empstudy}
	\end{center}
\end{figure}

{All the stations are located in a relatively small geographical region in Oklahoma State, and we excluded stations in the mountainous area in the northwestern part of the state. The largest distance between any two stations is about $531$ km, so we can reasonably assume that the data are stationary over space. We select the kernel function $k(\ss,\ss^*;\tht_K) = (1-\|\ss-\ss^*\|/r)_+^{\eta}$, $\tht_K=(\eta, r)^\top\in(0,\infty)^2$, and estimate its parameters using the inference approach described in Section~\ref{subsec-est-3} using close-by pairs at maximum distance $\delta_{\max}=300$\,km. Notice that although this kernel is compactly supported on $[0,r]$ with $r>0$ representing the dependence range, it reduces to the exponential kernel $k(\ss,\ss^*;\tht_K) = \exp(-\xi \|\ss-\ss^*\|)$ when $\eta:=\eta(r)$ such that $\eta(r)\to\infty$ and $\eta(r)/r \to \xi>0$ as $r \to \infty$. The estimated kernel parameters are $\widehat\tht_K =(\widehat\eta,\widehat r)^\top = (8985, 1270)^\top$ with the estimated range  $\widehat r>0$ expressed in units of $1000$ km. Thus, as expected, our results here imply fairly long range dependence for this temperature dataset, with $k(\ss,\ss^*) \approx e^{-7.08\|\ss-\ss^*\|}$; the fitted kernel function is plotted on the left panel of Figure~\ref{fig2empstudy}, from which the exponential decay is evident. Nevertheless, notice that our modeling approach has the great benefit of estimating whether the dependence range is finite or infinite (obtained as a boundary limiting case), rather than fixing a priori. Moreover, even if the range parameter $r$ is here much larger than the study region, the effective tail dependence range at which the estimated kernel function drops below $0.05$ is only about $423$km.} 

We then estimate the remaining parameters of the spatial mixture process~\eqref{eq-convgauss}, where the underlying Gaussian process is specified with a powered exponential correlation function defined as $\rho_G(\delta) = \{1 - \tau \mathbb I(\delta\neq0)\}\exp(-\theta_G \delta^{\alpha})$, $\delta\geq 0$, where $\mathbb I(\cdot)$ is the indicator function, $\theta_G>0$, $\alpha\in(0,2]$, and $\tau\in[0,1]$ is the nugget effect capturing small scale variability. More precisely, we here fix the kernel parameters to $\tht_K=(8985, 1270)^\top$ as estimated above, and then estimate the parameters $(\theta_G,\alpha,\tau,\beta)^\top$ using the least squares approach \eqref{eq:thtG-beta-LS} described in Section~\ref{subsec-est-3}. %Although we could in principle estimate the kernel parameters $\tht_K$ again based on \eqref{eq:thtK-LS2} as detailed in Section~\ref{subsec-est-3}, fixing them allows us to assess whether or not the fitted spatial mixture process provides an improvement over the reference Cauchy convolution model. 
We obtain $\widehat\theta_G=0.90\ (0.29)$, $\widehat\alpha = 0.82\ (0.20)$, $\widehat\tau = 0.013\ (0.018)$, and $\widehat \beta = 1.50\ (0.31)$, with standard errors calculated using the bootstrap shown in parentheses. %The fitted model thus corresponds to a Cauchy convolution model mixed with a Gaussian process. 
Since $\widehat\beta$ is positive and quite far from zero, the fitted spatial mixture process $\tilde{Z}(\ss)$ turns out to be quite different from the Cauchy convolution model $Z(\ss)$, providing increased flexibility to capture the behavior in the bulk of the distribution. Furthermore, the fitted process corresponds to a smooth Cauchy convolution process mixed with a quite rough Gaussian random field (with smoothness parameter $\widehat\alpha=0.82$), which yields realizations that are relatively---but not overly---smooth and thus more realistic. However, notice that the estimated nugget effect $\widehat\tau$ is here very small, indicating negligible micro-scale variability.

To assess the goodness-of-fit for the spatial mixture process~\eqref{eq-convgauss}, we then compute empirical and model-based Spearman's correlation estimates $S_{\rho}(\delta)$ for all pairs of sites $\{\ss_j,\ss_k\}$ and plot them in the middle panel of Figure~\ref{fig2empstudy} as a function of distance $\delta=\|\ss_j-\ss_k\|$. Model-based estimates are obtained by Monte Carlo using a large number of simulations from the fitted model. Similarly, the right panel of Figure~\ref{fig2empstudy} shows empirical and model-based estimates of the upper tail dependence coefficient $\lambda(\delta)$ plotted as function of distance $\delta$. We obtain very similar results for the lower tail dependence coefficient (not shown) as the data show no significant tail asymmetry. These plots show that the model \eqref{eq-convgauss} fits the data very well, both the in the bulk of the distribution (as measured by $S_{\rho}(\delta)$) and in the tails (as measured by $\lambda(\delta)$).

\begin{figure}[t!]
	\begin{center}
		\includegraphics[width=0.33\linewidth]{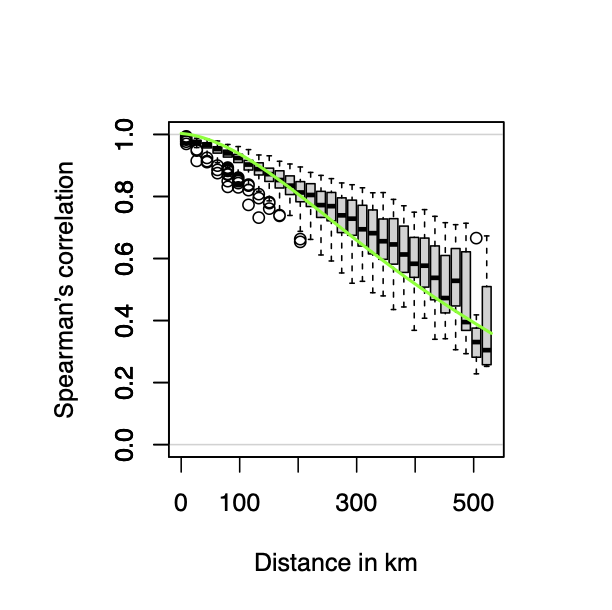}\hspace{-13mm}
		\includegraphics[width=0.33\linewidth]{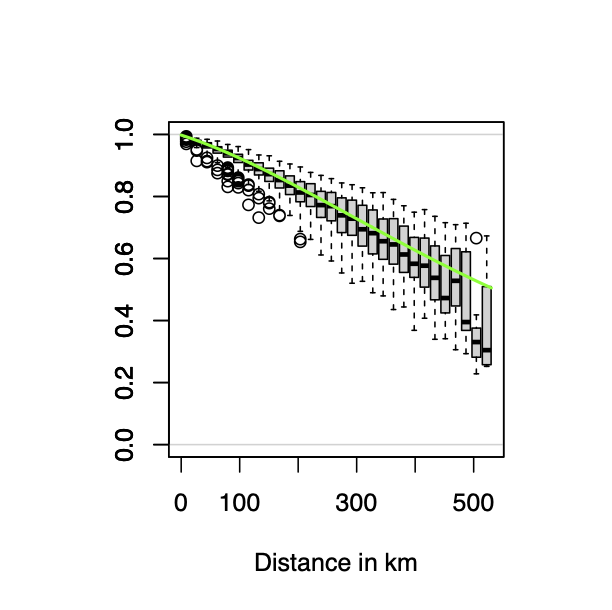}\hspace{-13mm}
		\includegraphics[width=0.33\linewidth]{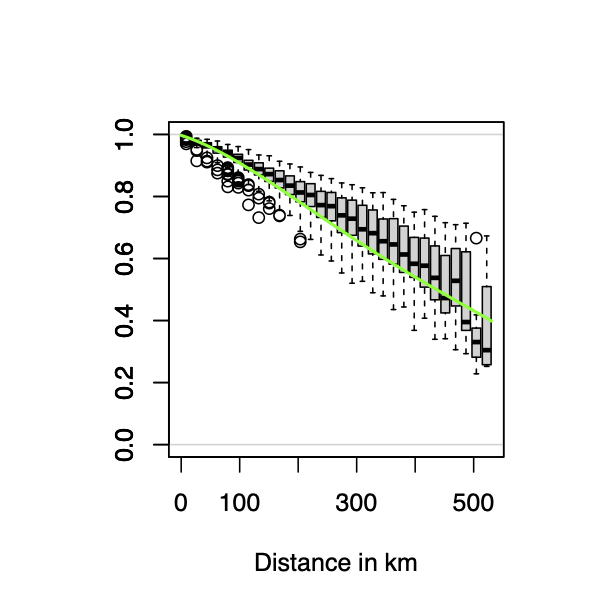}\vspace{-7mm}
		\includegraphics[width=0.33\linewidth]{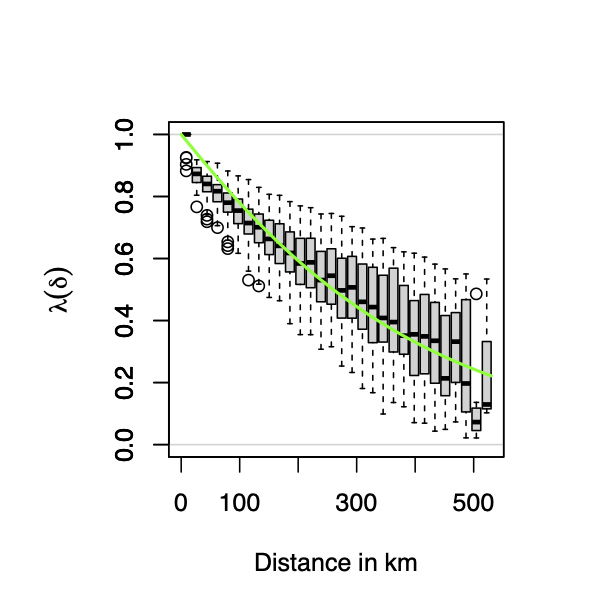}\hspace{-13mm}
		\includegraphics[width=0.33\linewidth]{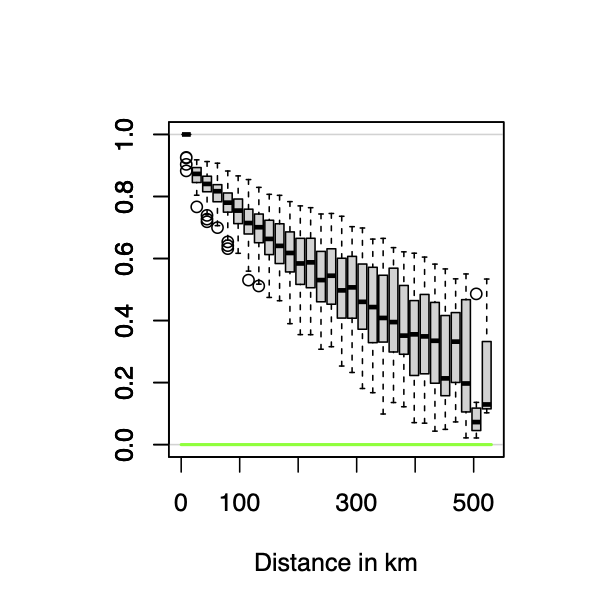}\hspace{-13mm}
		\includegraphics[width=0.33\linewidth]{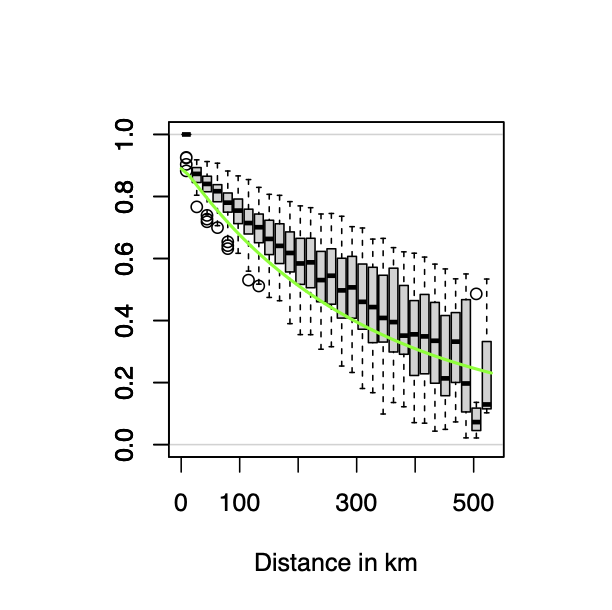}
		\caption{{\footnotesize { \emph{Top:} Empirical estimates of Spearman's correlation for all pairs of stations (boxplots) and fitted Spearman's correlation function $S_\rho(\delta)$ (solid green light), plotted as a function of distance $\delta$ between stations. \emph{Bottom:} Empirical upper tail dependence coefficient estimates for all pairs of stations (boxplots) and fitted tail dependence function $\lambda(\delta)$ (solid green light), plotted as a function of distance $\delta$ between stations. Estimates are obtained using Model 1 (left), Model 2 (middle) and Model 3 (right).}}}
		\label{fig3empstudy}
	\end{center}
\end{figure}

For comparison, we also fit some alternative copula models to the same temperature dataset. We consider the following models: the copula stemming from the Cauchy convolution process in \eqref{eq-convmodel} (Model~1); the Gaussian copula (Model~2) with the same powered-exponential correlation function as above for the process $Z_G$; and the Student-$t$ copula (Model~3) with $\nu>0$ degrees of freedom and same underlying correlation function as $Z_G$ and Model~2. Notice that Model~1 is a special case of our proposed model~\eqref{eq-convgauss} when $\beta=0$, and Model~2 can also be obtained as a limiting case of \eqref{eq-convgauss} as $\beta\to\infty$. We then compute empirical and model-based Spearman's correlation estimates $S_{\rho}(\delta)$ and the upper tail dependence coefficient $\lambda(\delta)$ for all pairs of sites $\{\ss_j,\ss_k\}$ and plot them in the first and second row of Figure~\ref{fig3empstudy} as a function of distance $\delta=\|\ss_j-\ss_k\|$. Model-based estimates are obtained by Monte Carlo using a large number of simulations from the fitted model.

Model~1 has a fairly good fit to the data, though the more flexible mixture process \eqref{eq-convgauss} has a slightly better fit in the bulk of the distribution, especially at relatively large distances. Model~2 has a good fit in the bulk of the distribution but cannot capture limiting tail dependence, and Model~3 tends to slightly underestimate the upper tail dependence structure, and is to rigid to separately control bulk and tail dependence characteristics. Overall, the spatial mixture process thus provides a better fit to the temperature data. We also recall here that our proposed mixture model \eqref{eq-convgauss} is the only one among the different models fitted that can capture short-range tail dependence and long-range tail independence, such that the proposed model should provide good fits on larger domains, as well.

\section{Concluding Remarks}
\label{sec-conc}

In this paper, we have proposed a Cauchy kernel convolution copula model for non-Gaussian spatial data and studied its dependence properties, with a particular eye on its tail behavior. In particular, with compactly supported kernels, our proposed model can capture complex dependence structures that possess short-range tail dependence, and long-range independence. Moreover, to further increase its flexibility in the bulk of the distribution and better capture the sub-asymptotic dependence behavior, we have also proposed a parsimonious copula model constructed by mixing a Cauchy convolution process with a Gaussian process. With this spatial mixture model extension, bulk and tail properties can be separately controlled with a few parameters, and a smooth transition between tail dependence classes is achieved as a function of distance between stations.

Our proposed inference scheme relies on a least-square minimum distance approach, which matches suitably chosen empirical and model-based summary statistics. It yields consistent estimates and it is guaranteed to be very fast even in high dimensions, thus circumventing the computational difficulties of likelihood-based inference. We have shown that our inference scheme works well using an extensive simulation study, and we have successfully applied it to a real temperature data example. Model parameters are generally easy to identify from each other, and the underlying kernel function can be accurately estimated, even in low sample sizes.

A limitation of our approach is that, by construction, the proposed Cauchy model and its spatial mixture extension are tail symmetric, and can only capture smooth extreme-value dependence structures characterized by moving maximum processes similar to the well-known Smith max-stable process \citep{Smith1990}. However, as opposed to fitting the (rigid) Smith model directly to spatially-indexed block maxima, we here propose to fit the Cauchy convolution process or the flexible spatial mixture process to the whole dataset instead. Therefore, even if the limiting extreme-value dependence structure is very smooth, our proposed models, which have much rougher realizations, can still adequately capture dependence characteristics at finite levels. 

Nevertheless, the sub-asymptotic tail structure of the spatial mixture extension of the Cauchy model is still quite restrictive in the sense that there is a discontinuity in the coefficient of tail dependence $\eta$ (reciprocal of the tail order) as a function of distance, being equal to
$\eta = 1$ for short distances and jumping to $\eta = 1/2$ at sufficiently large distances. The reason is that in the sum-mixture process $Z(\ss) + \beta Z_G(\ss)$, the heavy-tailed Cauchy convolution process $Z(\ss)$ dominates the Gaussian process $Z_G(\ss)$ in the tail. Unreported proofs suggest that this would remain the case if the margins of the $Z_G(\ss)$ process were replaced with a distribution with Weibull-like tails (i.e., potentially heavier-tailed than the Gaussian distribution but still lighter-tailed than the power-law Cauchy tails). However, despite this relatively restrictive limiting structure in the tail-independent case, we have shown through various theoretical diagnostics and goodness-of-fit plots in the data application that our model still possesses high flexibility at finite levels, all the way from low to high quantiles, separately capturing a wide range of different bulk and tail dependencies as a function of distance. Other constructions (e.g., considering product-mixtures rather than sum-mixtures) might be helpful to capture a smoother sub-asymptotic tail behavior, though this is still unclear and we leave this as a topic for future research. 

Another important problem is conditional simulation, i.e., simulation of the Cauchy convolution process conditional on the observed values of this process at some locations or conditional on the value of a spatial aggregation functional. This is a difficult problem for non-Gaussian kernel convolution processes since the likelihood function is intractable. One way to tackle conditional simulation could be to use some version of rejection sampling given that approximate simulations from the proposed model can be performed very fast. While such a naive approach would be computationally inefficient when performing simulations conditional on fixed values at a moderate or large number of spatial locations, it could prove helpful when conditioning the value of a single aggregation functional (e.g., the sum or maximum across sites). Approaches based on exponential tilting and importance sampling \citep{Rached.etal:2016, Botev.LEcuyer:2017} could potentially be adapted to further enhance conditional simulation, especially when the conditioning event is a low-probability rare event.

Further interesting research directions include generalizing our modeling approach to capture tail asymmetry, e.g., by considering kernel convolution processes constructed from asymmetric stable noise. It would also be interesting to study how to modify our model to capture rougher tail dependence structures of Brown--Resnick type.

\appendix
	\section{Proofs}
	%\sdescription{Short description of Supplement A.}
	\begin{lemma}[Lower tail probability]\label{eq:lemma1}
			For $j=1,\ldots,d$, let $Y_j^* = \sum_{k=1}^K a_{k,j}(m) Y_{k}$, where $a_{k,j}(m) \geq 0$ and $Y_1, \ldots, Y_K \sim_{\text{i.i.d.}} \mathrm{Cauchy(1)}$. Let $c_j(m,n) > 0$ and $\gamma_k = \gamma_k(m,n) := \max_{j=1,\ldots,d}\left\{a_{k,j}(m)/c_j(m,n)\right\}$ and assume that $\gamma_k \leq c_0/ (nK)$ for any $m, k$ and some constant $c_0 > 0$. Then for any $m$ and large enough $n$,
			$$
			\left|\Pr\{Y_1^* \leq c_1(m,n), \ldots, Y_d^* \leq c_d(m,n)\} - \left(1 - {1\over\pi} \sum_{k=1}^K \gamma_k\right)\right| \leq {(d+1)c_0^2\over\pi^2n^2}.
			$$
		\end{lemma}
		\noindent
		{\emph{Proof:}	 
			For the standard Cauchy CDF $F_{\CC}$ and $z > 1$,  we have
			$$
			1 - {1\over \pi z} \leq F_{\CC}(z) \leq 1 - {1\over\pi z} + {1\over 3\pi z^3}.
			$$
			By stability of the Cauchy distribution, we find thus that, for each $j=1,\ldots,d$ and large enough $n$,
			\begin{eqnarray*}
				\delta_j(m,n) &=& \Pr\left\{\sum_{k=1}^K a_{k,j}(m)Y_k \leq c_j(m,n)\right\}\\
				&&- \Pr\left\{Y_1 \leq {c_j(m,n)\over a_{1,j}(m)}, \ldots, Y_K \leq {c_j(m,n)\over a_{K,j}(m)}\right\}\\
				&\leq& 1 - {\sum_{k=1}^Ka_{k,j}(m)\over\pi c_j(m,n)}\left(1-{c_0^2\over 3n^2}\right) - \prod_{k=1}^K\left\{1 - {a_{k,j}(m)\over \pi c_j(m,n)}\right\}\\
				&\leq& 1 - {\sum_{k=1}^Ka_{k,j}(m)\over\pi c_j(m,n)}\left(1-{c_0^2\over 3n^2}\right) - 1 + {\sum_{k=1}^Ka_{k,j}(m)\over\pi c_j(m,n)}\\ 
				&\leq& {c_0^3\over 3\pi n^3} \leq {c_0^2\over \pi^2n^2},\\
				\delta_j(m,n) &\geq& 1 - {\sum_{k=1}^Ka_{k,j}(m)\over\pi c_j(m,n)} - \prod_{k=1}^K\left\{1 - {a_{k,j}(m)\over\pi c_j(m,n)}\left(1 - {c_0^2\over3\pi n^2K^2}\right)\right\}\\
				&\geq& -{c_0^3\over 3\pi^2n^3K} - \left(1 - {c_0^2\over 3\pi n^2K^2}\right)^2\sum_{k<l} {a_{k,j}(m)a_{l,j}(m)\over \pi^2 c_j(m,n)^2}\\ &\geq& -{c_0^3\over 3\pi^2n^3K} - \left(1 - {c_0^2\over 3\pi n^2K^2}\right)^2{c_0^2\over 2\pi^2 n^2} \geq - {c_0^2\over \pi^2 n^2}.
			\end{eqnarray*}
			Therefore, with $p_{m,n} = \Pr\{Y_1^* \leq c_1(m,n), \ldots, Y_d^* \leq c_d(m,n)\}$,
			$$
			\delta(m,n) = \left|p_{m,n} - \Pr\left(Y_1 \leq {1\over \gamma_1}, \ldots, Y_d \leq {1\over \gamma_d}\right)\right|
			\leq \sum_{j=1}^d|\delta_j(m,n)| \leq {dc_0^2\over \pi^2n^2}.
			$$
			Let $p_K^*=\Pr\left(Y_1 \leq {1\over\gamma_1}, \cdots,  Y_K \leq {1\over\gamma_K}\right)$. Now,
			\begin{eqnarray*}
				p_K^*&=& \prod_{k=1}^K F_{\CC}\left({1\over\gamma_{k}}\right) 
				\leq \prod_{k=1}^K\left\{1 - {\gamma_k\over\pi}\left(1 - {c_0^2\over3n^2}\right)\right\}\\
				&\leq& 1 - {1\over\pi}\left(1 - {c_0^2\over3n^2}\right) \sum_{k=1}^K \gamma_k + {1\over\pi^2}\left(1 - {c_0^2\over3n^2}\right)^2 \sum_{k<l}^K \gamma_k\gamma_l\\
				&\leq& 1 - {1\over\pi} \sum_{k=1}^K \gamma_k + {c_0^3\over3\pi n^3} + {c_0^2\over2 \pi^2 n^2} \leq 1 - {1\over\pi} \sum_{k=1}^K \gamma_k + {c_0^2\over\pi^2 n^2},\\
				p_K^* &\geq& \prod_{k=1}^K\left(1 - {\gamma_k\over\pi}\right) \geq  1 - {1\over\pi} \sum_{k=1}^K \gamma_k.
			\end{eqnarray*}
			Hence, $\left|p_{m,n} - \left(1 - {1\over\pi} \sum_{k=1}^K \gamma_k\right)\right| \leq {(d+1)c_0^2/(\pi^2n^2)}$ as required. \hfill $\Box$

			%%%%%%%%%%%%%%%%%%%%%%%%%%%%%
			%{\begin{prop} \label{prop1-EVlim1} \rm
			%		Consider the Cauchy process convolution defined as in \eqref{eq-convmodel} of the main paper with $W(\d\ss^*)\sim_{\text{i.i.d.}}\text{Cauchy}(\d\ss^*)$. For any collection of sites $\ss_1,  \ldots, \ss_d \in \mathbb{R}^2$, we write $Z_1 = Z(\ss_1), \ldots, Z_d = Z(\ss_d)$. Assume that $k(\ss,\ss^*) \leq k_{\max} < \infty$ is a nonnegative bounded integrable kernel function. Let $\ell(w_1, \ldots, w_d): [0,\infty)^d \mapsto [0,\infty)$ be the stable tail dependence function of the random vector $(Z_1, \ldots, Z_d)^\top$, then
			%		$$
			%		\ell(w_1, \ldots, w_d) =  \int_{\mathbb R^2} \max_{j=1,\ldots,d}w_j\zeta(\ss_j, \ss^*)\d\ss^*, \qquad  \zeta(\ss_j, \ss) = {k(\ss_j,\ss)\over\int_{\mathbb R^2} k(\ss_j,\ss^*)\d\ss^*}.  
			%		$$
			%\end{prop}}
			
			%\noindent
			{\emph{Proof of Proposition \ref{prop1-EVlim1}:}	 
				Define the expanding, but increasingly dense rectangular grid $S_m \subset \mathbb{R}^2$ as 
				$$S_m = \{\ss_{k,l}^*\}_{k,l=1}^m =\left\{-\frac{(m-1)\delta_m}{2}, -\frac{(m-3)\delta_m}{2}, \cdots, \frac{(m-3)\delta_m}{2}, \frac{(m-1)\delta_m}{2}\right\}^2,$$ 
				with $\delta_m\to0$ such that $m\delta_m \to \infty$, as $m \to \infty$. This implies that $\int_{S_m} h(x) \d x \to \int_{\mathbb{R}^2} h(x)\d x$, $m \to \infty$, for any integrable function $h(x)$. Define
				\begin{equation}
				\label{eq-appx-spatprocess}
				Z_{j,m} = \delta_m^2\sum_{k,l=1}^m k(\ss_j, \ss_{k,l}^*) W_{k,l}, \quad W_{k,l} = W(\ss_{k,l}^*) \sim_{\text{i.i.d.}} \text{Cauchy}(1).
				\end{equation}
				Intuitively, for large $m$, $Z_{j,m}$ is a finite approximation of $Z_j = Z(\ss_j)$. Formally, we indeed have $(Z_{1,m}, \ldots, Z_{d,m})^\top \to_d (Z_1, \ldots, Z_d)^\top$ as $m \to \infty$, where ``$\to_d$'' denotes convergence in distribution. 
				It can be seen that (\ref{eq-appx-spatprocess}) is a linear factor model with $m^2$ independent and identically distributed common factors $W_{k,l}$ as considered in \citet{Krupskii.Genton2018}.}
			
			{Let $F_{j,m}$ be the CDF of $Z_{j,m}$, $j=1,\ldots,d$. Define 
				$$\ell_{m,n}(w_1, \ldots,w_d) = n\left[1 - \Pr\left\{Z_{1,m} \leq F^{-1}_{1,m}(1-\frac{w_1}{n}), \ldots, Z_{p,m} \leq F^{-1}_{d,m}(1-\frac{w_d}{n})\right\}\right].$$ 
				The stable tail dependence function associated with the random vector $(Z_{1,m}, \ldots, $ $Z_{d,m})^\top$ is 
				$$\ell_m(w_1, \ldots, w_d) = \lim_{n \to \infty}\ell_{m,n}(w_1, \ldots, w_d).$$ 
				By Moore-Osgood Theorem (see \citet{Angus2012}, p.~140), we need to prove that $\ell_{m,n}(w_1, \ldots, w_d)$ converges uniformly to $\ell_m(w_1, \ldots, w_d)$, as $n \to \infty$.}
			
			{By stability of the Cauchy distribution, we have $Z_{j,m} \sim \mathrm{Cauchy}(c_{j,m})$ with scale parameter $c_{j,m} =  \delta_m^2\sum_{k,l=1}^m k(\ss_j, \ss_{k,l}^*)$, and therefore 
				$$
				F_{j,m}^{-1}\left(1-{w_j\over n}\right) = c_{j,m}\cot\left({\pi w_j\over n}\right),\qquad j=1,\ldots,d.
				$$
				Using the Laurent series of the cotangent function around the origin with $\cot(z) = {1/z} - {z/3} + \epsilon(z)$, $|\epsilon(z)| < z^3$ for $z < 1$, we deduce that
				$$
				{nc_{j,m}\over\pi w_j}\left(1 - {c_L\over n^2}\right) \leq {nc_{j,m}\over\pi w_j} - {\pi w_jc_{j,m}\over2n} \leq F_{j,m}^{-1}\left(1-{w_j\over n}\right) \leq {nc_{j,m}\over\pi w_j} %\leq c_{U} n
				$$
				for any $m$ if $n$ is large enough, where $c_L = 2^{-1}\pi^2\max_{j=1,\ldots,d}w_j^2, \ %c_U = \max_{i=1,2} \{(\xi_i + k_{\max})/w_i\}
				$ is a constant that does not depend on $m$ and $n$. We now use Lemma~\ref{eq:lemma1}.} 
			Let $p_{m,n} = p_{m,n}(w_1, \ldots, w_d) = \Pr\{Z_{1,m} \leq \eta_{1,m}(w_1,n), \ldots, Z_{d,m} \leq \eta_{d,m}(w_d,n)\}$. Using Lemma~\ref{eq:lemma1} with $K = m^2$, $c_j(m,n) = F_{j,m}^{-1}\left(1-{w_j\over n}\right)$ and $c_0 = 2\pi\max_{\ss, j} \{w_j\zeta(\ss_j,\ss)\}$, we find for $\ell_{m,n}(w_1,\ldots,w_d) = n\{1-p_{m,n}(w_1,\ldots,w_d)\}$ that 
			\begin{eqnarray*}
				|\ell_{m,n}(w_1,\ldots,w_d) - \ell_{m}^*(w_1,\ldots,w_d)| &\leq& {(d+1)c_0^2\over \pi^2 n},\\
				|\ell_{m}^*(w_1,\ldots,w_d) - \ell_{m}(w_1,\ldots,w_d)| &\leq& {c_L\over (n^2-c_L)}\,\ell_m(w_1, \ldots, w_d)\\
				&\leq& {c_L\over (n^2-c_L)}\,M(w_1, \ldots, w_d),
			\end{eqnarray*}
			where $M(w_1, \ldots, w_d) = \max_m \ell_m(w_1, \ldots, w_d)$ and
			\begin{eqnarray*}
				\ell_m^*(w_1, \ldots, w_d) &=& {n\delta_m^2\over \pi}\sum_{k,l=1}^m \max_{j=1,\ldots,d}\left\{{k(\ss_j,\ss_{k,l}^*)\over F_{j,m}^{-1}\left(1-{w_j\over n}\right)}\right\},\\
				\ell_m(w_1, \ldots, w_d) &=& \delta_m^2\sum_{k,l=1}^m \max_{j=1,\ldots,d}\left\{{w_jk(\ss_j,\ss_{k,l}^*)\over c_{j,m}}\right\},
			\end{eqnarray*}
			It follows that $\ell_{m,n}(w_1, \ldots, w_d) \to \ell_m(w_1, \ldots, w_d)$, as $n \to \infty$, uniformly over $m$.}
		
		{This implies that the stable tail dependence function associated with the $d$-dimensional random vector $(Z_1, \ldots, Z_d)^\top$ can be calculated as 
			%\begin{align*}
			$\ell(w_1, \ldots, w_d) = \lim_{n\to \infty}\lim_{m \to \infty} \ell_{m,n}(w_1, \ldots, w_d) = \lim_{m\to \infty}\lim_{n \to \infty} \ell_{m,n}(w_1, \ldots, w_d) = $\\$\lim_{m \to \infty} \ell_{m}(w_1, \ldots, w_d) = \int_{\mathbb R^2} \max_{j=1,\ldots,d}w_j\zeta(\ss_j, \ss^*)\d\ss^*$, 
			%\end{align*} 
			and the result of the proposition follows. \hfill $\Box$}
		
		%%%%%%%%%%%%%%%%%%%%%%%%%%%%%
		%{\begin{prop} \label{prop1b-EVlim1} \rm
		%		Suppose that the assumptions of Proposition~\ref{prop1-EVlim1} hold. Moreover, assume that the kernel function in \eqref{eq-convmodel} of the main paper may be written as $k(\ss,\ss^*) = g(\|\ss-\ss^*\|)$, where $g$ is an integrable nonnegative monotonically decreasing function. Then, for any sites $\ss_1, \ss_2\in\mathbb R^2$, the stable tail dependence function $\ell(w_1,w_2)$ of $\{Z(\ss_1),Z(\ss_2)\}^\top$ satisfies $\ell(w_1, w_2) \leq \{1+K\delta + o(\delta)\}\max(w_1, w_2)$, as $\delta:=\|\ss_1-\ss_2\| \to 0$, where $K$ is some constant that does not depend on $w_1$ and $w_2$. Furthermore, we can select $K$ such that $\ell(1,1) = 1+K\delta + o(\delta)$.
		%\end{prop}}
		
		%\noindent
		{\emph{Proof of Proposition \ref{prop1b-EVlim1}:} We first prove the second part of the proposition. Without loss of generality, we assume that $\int_{\mathbb R^2}g(\|\ss^*\|)\d \ss^*=1$. Using stationarity of the kernel and an orthonormal transform of the integration variables, we obtain from Proposition~\ref{prop1-EVlim1} that
			\begin{align*}
			\ell(1,1) &= \int_{\mathbb R^2} \max_{j=1,2} g(\|\ss_j - \ss^*\|) \d\ss^* \\
			&= \int_{\mathbb R^2} \max \{g(\|(s^*_1, s^*_2)^\top\|), g(\|(s^*_1 + \delta, s^*_2)^\top\|)\} \d\ss^*,\qquad \ss^*=(s_1^*,s_2^*)^\top,\\
			&= 1 + \int_{\{\ss^*\in\mathbb R^2:s_1^* < -\delta/2\}} \{g(\|(s^*_1 + \delta, s^*_2)^\top\|) - g(\|(s^*_1, s^*_2)^\top\|)\}\d\ss^*\\
			&= 1 + \int_{-\delta/2}^{\delta/2}\int_{\mathbb{R}} g(\|(s^*_1, s^*_2)^\top\|) \d s^*_2 \d s^*_1 = 1+\delta\int_{\mathbb{R}} g(\|(0, s^*_2)^\top\|)\d s^*_2 + o(\delta),
			\end{align*}
			so one can take $K = \int_{\mathbb{R}} g(\|(0, s^*_2)^\top\|)\d s^*_2$.}
		
		{Now we prove the first part of the proposition.  Without loss of generality, we assume that $w_1 > w_2$. We get $\ell(w_1,w_2) - w_1 = \int_{S(w_1,w_2)} \{w_2 g(\|(s_1^*+\delta,s_2^*)^\top\|) - w_1g(\|(s_1^*,s_2^*)^\top\|)\} \d\ss^*$, where $S(w_1,w_2) = \{\ss^*\in\mathbb R^2: w_2 g(\|(s_1^*+\delta,s_2^*)^\top\|) > w_1g(\|(s_1^*,s_2^*)^\top\|)\} \subset \{\ss^*\in\mathbb R^2: s_1^* < - \delta/2\}$, so
			\begin{align*}
			\ell(w_1,w_2) - w_1 &\leq w_2\int_{S(w_1,w_2)} \{g(\|(s_1^*+\delta,s_2^*)^\top\|) - g(\|(s_1^*,s_2^*)^\top\|)\} \d\ss^*\\
			&\leq  w_2\int_{\{\ss^*\in\mathbb R^2:s_2^* < -\delta/2\}} \{g(\|(s_1^*+\delta,s_2^*)^\top\|) - g(\|(s_1^*,s_2^*)^\top\|)\} \d\ss^*\\ 
			&= w_2\{K\delta  + o(\delta)\} \leq  w_1\{K\delta+o(\delta)\},
			\end{align*}
			where the last equality follows the same line of proof as above. This implies that $\ell(w_1,w_2) \leq \{1+K\delta + o(\delta)\}w_1$, which concludes the proof of the proposition.  \hfill $\Box$}
		
		%%%%%%%%%%%%%%%%%%%%%%%%%%%%%
		%{\begin{corol} \label{corol} \rm Under the assumptions of Proposition \ref{prop1b-EVlim1}, the limiting extreme-value process $Z_{\rm EV}(\ss)$ of the Cauchy convolution process $Z(\ss)$ (defined as in \eqref{eq-convmodel} of the main paper with $W(\d\ss^*)\sim_{\text{i.i.d.}}\text{Cauchy}(\d\ss^*)$) has dependence coefficients satisfying $\lambda(\delta) = 1 - K\delta + o(\delta)$ and $S_{\rho}(\delta) > 1 - 4K\delta$, as $\delta \to 0$. 
		%\end{corol}}
		
		%\noindent
		{\emph{Proof of Corollary \ref{corol}:} It is easy to verify that $\lambda(\delta) = 2 - \ell(1,1)$, where $\ell$ is the stable tail dependence function for two sites $\ss_1,\ss_2\in\mathbb R^2$ at distance $\delta=\|\ss_1-\ss_2\|$ from each other, and by Proposition~\ref{prop1b-EVlim1}, we therefore obtain $\lambda(\delta) = 1 - K\delta + o(\delta)$. Moreover, as the function $\ell(w_1,w_2)$ in non-decreasing in both arguments and homogeneous of order $1$, we get $\ell(-\log u_1,-\log u_2) \leq -\log\{\min(u_1,u_2)\}\ell(1,1)=\log\{\min(u_1,u_2)\}\{1+K\delta+o(\delta)\}$ thanks to Proposition~\ref{prop1b-EVlim1}. This yields
			\begin{align*}
			S_{\rho}(\delta) &= 12\iint_{[0,1]^2} C_{\rm EV}(u_1,u_2) \d u_1 \d u_2 - 3 \\
			&= 12\iint_{[0,1]^2} \exp\{-\ell(-\log u_1, -\log u_2)\} \d u_1 \d u_2 - 3\\
			&\geq 12\iint_{[0,1]^2} \min(u_1,u_2)^{1+K\delta}\d u_1 \d u_2 + o(\delta) - 3\\
			&= 12 {2\over 6+5K\delta+o(\delta)}-3 + o(\delta) \\
			&= {6-15K\delta+o(\delta)\over 6+5K\delta+o(\delta)}+ o(\delta) = 1 - {10K\delta\over 3} + o(\delta) > 1 - 4K\delta, \quad\delta \to 0,
			\end{align*}
			which concludes the proof. \hfill $\Box$}

		%%%%%%%%%%%%%%%%%%%%%%%%%%%%%
		%{\begin{corol} \label{corol2} \rm Assume that the assumptions of Proposition \ref{prop1b-EVlim1} hold, and that 
		%		$$G(\delta) = \max_{(s_1^*,s_2^*)^\top\in\mathcal{S}_{\cup+}(\delta)} g(\|(s_1^* + \delta, s_2^*)^\top\|)/g(\|(s_1^*, s_2^*)^\top\|)  < \infty,$$
		%		where $\mathcal{S}_{\cup+}(\delta):=\{(s_1^*,s_2^*)\in\mathbb R^2:g(\|(s_1^*, s_2^*)^\top\|)>0\text{ and/or } g(\|(s_1^* + \delta, s_2^*)^\top\|)>0\}$, with the convention that $x/0=\infty$ for all $x>0$. Then, $\ell(w_1, w_2) = w_1$ for all $(w_1,w_2)^\top\in[0,\infty)^2$ with $w_1/w_2 > G(\delta)\geq1$.  Similarly,  $\ell(w_1, w_2) = w_2$ for all $(w_1,w_2)^\top\in[0,\infty)^2$ with $w_2/w_1 > G(\delta)\geq1$.
		%\end{corol}}
		
		%\noindent
		{\emph{Proof of Corollary \ref{corol2}:} If $w_1/w_2 > G(\delta)$, then $S(w_1, w_2) = \{(s_1^*,s_2^*)^\top\in\mathbb R^2: w_2 g(\|(s_1^*+\delta,s_2^*)^\top\|) > w_1g(\|(s_1^*,s_2^*)^\top\|)\}$ is the empty set, i.e., $S(w_1,w_2)=\emptyset$, which implies that $\ell(w_1,w_2) = w_1$ by following the proof of Proposition~\ref{prop1b-EVlim1}. The second part of the corollary follows by symmetry. \hfill $\Box$}

		%%%%%%%%%%%%%%%%%%%%%%%%%%%%%
		%{\begin{prop} \label{prop2-EVlim1} \rm
		%		Under the assumptions of Proposition \ref{prop1-EVlim1}, the stable tail dependence function corresponding to the joint distribution of $\{\tilde{Z}(\ss_1), \ldots, \tilde{Z}(\ss_d)\}^\top$, with the process $\tilde{Z}(\ss)$ defined as in (7) of the main paper, is $\ell(w_1, \ldots, w_d)$ given in Proposition~\ref{prop1-EVlim1}.  	
		%\end{prop}}
		
		%\noindent
		{\emph{Proof of Proposition \ref{prop2-EVlim1}:} Without loss of generality, we assume that \\ $\int_{\mathbb{R}^2}
			k(\ss, \ss^*)\d\ss^* = 1$ for any $\ss\in\mathbb R^2$. Let $F^{-1}_{\CC}(\cdot)$ be the quantile function of the standard Cauchy distribution. Denote $Z_j = Z(\ss_j)$, $Z_{G,j} = Z_G(\ss_j)$, $\tilde{Z}_{j} = \tilde{Z}(\ss_j)$, and $h_j = h_j(w_j,n) = F_{\CC}^{-1}(1 - w_j/n)$, $j=1,\ldots,d$. We have:
			$$
			\Pr(\tilde{Z}_1 \leq h_1, \ldots, \tilde{Z}_d \leq h_d) = \Pr(\tilde{Z}_1 \leq h_1, \ldots, \tilde{Z}_d \leq h_d, \max_{j=1,\ldots,d} |\beta Z_{G,j}| \leq \sqrt{n}) + \epsilon_n,
			$$
			where, thanks to Mills's ratio, \begin{align*}
			0 < \epsilon_n &\leq \Pr(\max_{j=1,\ldots,d} |\beta Z_{G,j}| > \sqrt{n}) < \sum_{j=1}^d \Pr(|\beta Z_{G,j}| > \sqrt{n}) \\
			&\leq d |\beta|\phi(\sqrt{n}/|\beta|)/\sqrt{n} = o(e^{-n/(2\beta^2)}),\end{align*}
		    and $\phi(\cdot)$ denotes the density function of the standard normal distribution.  Let 
			$$p^* = p^*(w_1,\ldots,w_d,n) = \Pr(\tilde{Z}_1 \leq h_1, \ldots, \tilde{Z}_d \leq h_d, \max_{j=1,\ldots,d} |\beta Z_{G,j}| \leq \sqrt{n}).$$
			We find that 
			\begin{align*}
			p^* &\leq \Pr(Z_1 \leq h_1 + \sqrt{n}, \ldots, Z_d \leq h_d + \sqrt{n})= \Pr(Z_1 \leq h_1, \ldots, Z_d \leq h_d) + o(n^{-1}),\\
			p^* &\geq  \Pr(Z_1 \leq h_1 - \sqrt{n}, \ldots, Z_d \leq h_d - \sqrt{n}, \max_{j=1,\ldots,d}|\beta Z_{G,j}| \leq \sqrt{n})\\ 
			&= \Pr(Z_1 \leq h_1 - \sqrt{n}, \ldots, Z_d \leq h_d - \sqrt{n}) + o(e^{-n/(2\beta^2)})\\
			&= \Pr(Z_1 \leq h_1, \ldots, Z_d \leq h_d) + o(n^{-1}).
			\end{align*}
			Let $\tilde{h}_j(w_j, n)$ be the $(1-w_j/n)$-quantile of $\tilde{Z}_j$, $j=1,\ldots, d$. Note that $\Pr(\tilde Z_j < h_j) = \Pr(Z_j < h_j) + o(n^{-1})$, so $\tilde{h}_j(w_j, n) = h_j(w_j, n) + o(n)$, and therefore
			\begin{align*}
			\Pr(\tilde{Z}_1 \leq \tilde{h}_1, \ldots, \tilde{Z}_d \leq \tilde{h}_d) &= \Pr(\tilde{Z}_1 \leq h_1, \ldots, \tilde{Z}_d \leq h_d) + o(n^{-1})\\
			&= \Pr(Z_1 \leq h_1, \ldots, Z_d \leq h_d) + o(n^{-1}),
			\end{align*}
			and $(Z_1, \ldots, Z_d)^\top$ and $(\tilde{Z}_1, \ldots, \tilde{Z}_d)^\top$ thus have the same stable tail dependence function. \hfill $\Box$}
		
\section{Special cases}

\begin{example}[Smith model \citep{Smith1990} in $\mathbb{R}^2$]
	{Assume that $q=2$ and that $k(\ss,\ss^*) = \phi_{2}(\ss-\ss^*;\sigma,\rho)$ where $\phi_{2}(\cdot;\sigma,\rho)$ denotes the density function the bivariate Gaussian distribution with zero mean, variances both equal to $\sigma^2$, and correlation $\rho\in(-1,1)$. It follows that  
		\begin{eqnarray*}
			\ell(w_1,w_2) &=& w_1\int_{{\phi_2(\ss*+\Delta\ss;\sigma,\rho)\over\phi_2(\ss^*;\sigma,\rho)} \leq {w_1\over w_2}} \phi_2(\ss^*;\sigma,\rho)\d\ss^* \\
			&& + \, w_2\int_{{\phi_2(\ss*-\Delta\ss;\sigma,\rho)\over\phi_2(\ss^*;\sigma,\rho)} \leq {w_2\over w_1}} \phi_2(\ss^*;\sigma,\rho)\d\ss^*\\
			&=& w_1\Phi\left({\lambda^*_2\over 2} + {1\over\lambda^*_2}\log{w_1\over w_2}\right) + w_2\Phi\left({\lambda^*_2\over 2} + {1\over\lambda^*_2}\log{w_1\over w_2}\right),
		\end{eqnarray*}
		where $\Delta\ss = \ss_1 - \ss_2$ and $\lambda^*_2 = \{\gamma(\Delta\ss)\}^{1/2}$ with variogram $\gamma(\mathbf{h}) = \sigma^{-2}\mathbf{h}^\top\left(\begin{matrix}
		1 & \rho \\
		\rho & 1 \\
		\end{matrix}\right)^{-1}\mathbf{h}$. 
		When $\rho=0$, we obtain the isotropic variogram $\gamma(\mathbf{h})=(\|\mathbf{h}\|/\sigma)^2$, which is similar to the case in $\mathbb R$ above. Similar expressions can be obtained when variances are not equal (see \citet{Smith1990}) or when the kernel's variance-covariance matrix varies spatially (see \citet{Huser.Genton:2016}).}
\end{example}

\begin{example}[Kernels with compact support]
	{As another example with compactly supported kernel, consider the truncated Gaussian kernel $k(s,s^*) = {1\over A}\phi(s-s^*; \sigma^2)\ii\{|s-s^*| < r\}$ where $\phi(\cdot; \sigma^2)$ is Gaussian density with mean zero and variance $\sigma^2$, and $A = 2\Phi(r/\sigma) - 1$ is the normalizing factor such that $\int_{\mathbb R}k(s,s^*)\d s^*=1$. Again, for simplicity we assume $s_1 < s_2$. If $\delta=s_2 - s_1 \geq 2r$, then $\ell(w_1, w_2) = w_1 + w_2$. Otherwise,  
		\begin{align*}
		M(s_1, s_2) &= \{s^*\in\mathbb R: w_1\zeta(s_1, s^*) > w_2\zeta(s_2, s^*)\}\\ &= \left\{s^*\in\mathbb R: s_1 - r < s^* < s_2 - r\right\},
		\end{align*} 
		for $\log(w_1/w_2) \geq -\Delta$, where $\Delta = \delta(r - \delta/2)/\sigma^2$, and 
		$$M(s_1, s_2) = \left\{s^*\in\mathbb R: s_1 - r < s^* < s_1 + \min\left(r, {\sigma^2\over\delta}\log{w_1\over w_2} + {\delta\over2}\right)\right\},$$ for $\log(w_1/w_2) \leq - \Delta$. This implies that
		$$
		\ell(w_1,w_2) = \begin{cases}
		{w_1\over A}\left\{\Phi({\sigma\over\delta}\log{w_1\over w_2} + {\delta\over2\sigma}) - \Phi(-{r\over\sigma})\right\} & \\
		\hspace{2cm} + {w_2\over A}\left\{\Phi({\sigma\over\delta}\log{w_2\over w_1} + {\delta\over2\sigma}) - \Phi(-{r\over\sigma})\right\}, & \left|\log{w_1\over w_2}\right| \leq \Delta,\\
		{w_1\over A}\left\{\Phi({\delta-r\over\sigma}) - \Phi(-{r\over\sigma})\right\} + w_2, &  \log{w_1\over w_2} < -\Delta,\\
		w_1+{w_2\over A}\left\{\Phi({\delta-r\over\sigma}) - \Phi(-{r\over\sigma})\right\}, &  \log{w_1\over w_2} > \Delta.
		\end{cases}
		$$}
\end{example}

%%%%%%%%%%%%%%%%%%%%%%%%%%%%%%%%%%%%%%%%%%%%

\section{Approximate simulation of the max-stable process \eqref{eq:movingmaximum}}

{The next proposition can be used to easily and quickly generate approximate simulations from the limiting moving maximum max-stable process corresponding to the process $Z(\ss)$ (or $\tilde{Z}(\ss))$ with arbitrary accuracy.}

{\begin{prop}[Approximate simulation of $Z_{\rm EV}$ on a dense grid] \label{prop1-EVlim2} \rm
		Let $Z_{\rm EV}(\ss)$ be the moving maximum max-stable process \eqref{eq:movingmaximum} of the main paper with unit Fr\'echet margins $F^*(z) = \exp(-1/z)$, $z>0$, which corresponds to extreme-value limit of the Cauchy convolution process $Z(\ss)$ defined in \eqref{eq-convmodel} of the main paper with $W(\d\ss^*)\sim_{\text{i.i.d.}}\text{Cauchy}(\d\ss^*)$. Moreover, let $\{\ss_{k,l}\}_{k,l=1}^m$ be the rectangular grid $S_m \subset \mathbb{R}^2$ as defined in the proof of Proposition~\ref{prop1-EVlim1} with the cell size $\delta_m \times \delta_m$ with $\delta_m \to 0$ such that $m\delta_m\to\infty$, as $m\to \infty$.  Following the notation of Proposition~\ref{prop1-EVlim1}, define 
		$$
		Z_{{\rm EV};j,m} = \delta_m^2 \max_{k,l=1,\ldots,m} \left\{\zeta(\ss_j, \ss_{k,l}) W_{k,l}^*\right\}, \quad W_{k,l}^* \sim_{\text{i.i.d}} F^*,\quad j=1,\ldots,d.
		$$
		It follows that $(Z_{{\rm EV};1,m}, \ldots, Z_{{\rm EV};d,m})^\top \to_d \{Z_{\rm EV}(\ss_1), \ldots, Z_{\rm EV}(\ss_d)\}^\top$, as $m \to \infty$.
\end{prop} }

\noindent 
{\emph{Proof:} Define 
	%\begin{equation*}
	$Z_{j,m} = \delta_m^2\sum_{k,l=1}^m \zeta(\ss_j,\ss_{k,l}) W_{k,l}, \quad W_{k,l} = W(\ss_{k,l}) \sim_{\text{i.i.d.}} \text{\text{Cauchy}}(\pi)$. 
	%\end{equation*}
	Without loss of generality we use i.i.d. Cauchy random variables with the scale parameter $\pi$ here so that $\Pr(W_{k,l} \leq z) = 1 - 1/z + o(1/z)$ and $\Pr(W_{k,l}^* \leq z) = 1 - 1/z + o(1/z)$ as $z \to \infty$.} 

{Let $\tilde q_{m,j,h}$ and $q_{m,j,h}$ be the $(1-h/n)$-quantiles of $Z_{{\rm EV};j,m}$ and $Z_{j,m}$, respectively. We have $\tilde q_{m,j,h} - q_{m,j,h} = o(n)$ (see Chapter VIII.8  in \citet{Feller1970}), and therefore, with $\tilde p_{m,d} = \Pr\{Z_{\rm EV;1,m} \leq \tilde q_{m,1,h_1}, \ldots, Z_{{\rm EV};d,m}  \leq \tilde q_{m,d,h_d}\}$,
	\begin{eqnarray*}
		 \tilde p_{m,d} &=& \prod_{k,l=1}^m\Pr\left\{W_{k,l}^* \leq \min_{j=1,\ldots,d}{m^{2}\tilde q_{m,j,h_j}\over\zeta(\ss_j,\ss_{k,l})}\right\}\\ 
		&=& \prod_{k,l=1}^m\Pr\left\{W_{k,l}^* \leq \min_{j=1,\ldots,d}{m^{2}q_{m,j,h_j}\over\zeta(\ss_j,\ss_{k,l})}\right\} + o\left(n^{-1}\right).
	\end{eqnarray*}
	Let $p_{m,d} = 	\Pr\{Z_{1,m} \leq q_{m,1,h_1}, \ldots, Z_{d,m} \leq q_{m,d,h_d}\}$. Using Lemma~\ref{eq:lemma1}, 
	\begin{eqnarray*}
	p_{m,d} &=& \prod_{k,l=1}^m\Pr\left\{W_{k,l} \leq \min_{j=1,\ldots,d}{m^{2}q_{m,j,h_j}\over\zeta(\ss_j,\ss_{k,l})}\right\} + o\left(n^{-1}\right)\\
		&=& \prod_{k,l=1}^m\Pr\left\{W_{k,l}^* \leq \min_{j=1,\ldots,d}{m^{2}q_{m,j,h_j}\over\zeta(\ss_j,\ss_{k,l})}\right\} + o\left(n^{-1}\right).
	\end{eqnarray*}
	This implies that the limiting extreme-value distributions of $\ZZ_m = (Z_{1,m},\ldots,$ $Z_{d,m})^\top$ and $\tilde\ZZ_m = (Z_{{\rm EV};1,m},\ldots,Z_{{\rm EV};d,m})^\top$ are the same. By Proposition 1, we obtain that $(Z_{{\rm EV};1,m}, \ldots, Z_{{\rm EV};d,m})^\top \to_d \{Z_{\rm EV}(\ss_1), \ldots, Z_{\rm EV}(\ss_d)\}^\top$ as $m \to \infty$. \hfill $\Box$}

%{\begin{prop} \label{prop1-EVlim2} \rm
%		Let $Z_{\rm EV}(\ss)$ be the moving maximum max-stable process \eqref{eq:movingmaximum} with unit Fr\'echet margins $F^*(z) = \exp(-1/z)$, $z>0$, which corresponds to extreme-value limit of the Cauchy convolution process $Z(\ss)$ defined in \eqref{eq-convmodel} with $W(\d\ss^*)\sim_{\text{i.i.d.}}\text{Cauchy}(\d\ss^*)$. Moreover, let $\{\ss_{k,l}\}_{k,l=1}^m$ be the rectangular grid $S_m \subset \mathbb{R}^2$ as defined in the proof of Proposition~\ref{prop1-EVlim1} with the cell size $\delta_m \times \delta_m$ with $\delta_m \to 0$ such that $m\delta_m\to\infty$, as $m\to \infty$.  Following the notation of Proposition~\ref{prop1-EVlim1}, define 
%		$$
%		Z_{{\rm EV};j,m} = m^{-2} \max_{k,l=1,\ldots,m} \left\{\zeta(\ss_j, \ss_{k,l}) W_{k,l}^*\right\}, \quad W_{k,l}^* \sim_{\text{i.i.d}} F^*,\quad j=1,\ldots,d.
%		$$
%		It follows that $(Z_{{\rm EV};1,m}, \ldots, Z_{{\rm EV};d,m})^\top \to_d \{Z_{\rm EV}(\ss_1), \ldots, Z_{\rm EV}(\ss_d)\}^\top$, as $m \to \infty$.
%\end{prop} }

%%%%%%%%%%%%%%%%%%%%%%%%%%%%%%%%%%%%%%%%%%%%%%%%
%%%%%%%%%%%%%%%%%%%%%%%%%%%%%%%%%%%%%%%%%%%%%%%%
%%%%%%%%%%%%%%%%%%%%%%%%%%%%%%%%%%%%%%%%%%%%%%%%

\bibliographystyle{model2-names}
\bibliography{ref1}

\end{document}